\documentclass[journal]{IEEEtran}
\usepackage{graphicx}
\usepackage{amsmath}
\usepackage{multirow}
\usepackage{float}
\usepackage{bm}
\usepackage{amsfonts}
\usepackage{cite}
\usepackage{hyperref}

\ifCLASSINFOpdf
\else
\fi
\usepackage{cite}
\usepackage[numbers,sort&compress]{natbib}
\begin{document}
%

\title{CampNet: Context-Aware Mask Prediction for End-to-End Text-Based Speech Editing}
%
%

\author{
Tao~Wang,
Jiangyan~Yi,~\IEEEmembership{Member,~IEEE,}
   Ruibo~Fu,~\IEEEmembership{Member,~IEEE,}
      Jianhua~Tao,~\IEEEmembership{Senior~Member,~IEEE,}
        and~Zhengqi~Wen
\thanks{T. Wang is with the National Laboratory of Pattern Recognition, Institute of
Automation, Chinese Academy of Science, Beijing 100190, China, and also
with the School of Artificial Intelligence, University of Chinese
Academy of Sciences, Beijing 100190, China (e-mail: tao.wang@nlpr.
ia.ac.cn).}
\thanks{ J. Yi, R. Fu, J. Tao and Z. Wen are  with the National Laboratory of Pattern Recognition,
Institute of Automation, Chinese Academy of Sciences, Beijing 100190, China
(e-mail: \{jiangyan.yi, ruibo.fu,  jhtao, zqwen\}@nlpr.ia.ac.cn).}

\thanks{Corresponding Author: Jiangyan Yi, Ruibo Fu,    Jianhua Tao.  E-mail: \{jiangyan.yi, ruibo.fu, jhtao\}@nlpr.ia.ac.cn.}}

%
%

\markboth{Journal of \LaTeX\ Class Files,~Vol.~14, No.~8, August~2015}%
{Shell \MakeLowercase{\textit{et al.}}: Bare Demo of IEEEtran.cls for IEEE Journals}
%



\maketitle

\begin{abstract}
The text-based speech editor allows the editing of speech through intuitive cutting, copying, and pasting operations to speed up the process of editing speech. However, the major drawback of current systems is that edited speech often sounds unnatural due to cut-copy-paste operation. In addition, it is not obvious how to synthesize records according to a new word not appearing in the transcript, which often needs the help of text-to-speech (TTS) and voice conversion (VC) technology at the same time. This paper first proposes a novel end-to-end text-based speech editing method called context-aware mask prediction network (CampNet). The model can simulate the text-based speech editing process by randomly masking part of speech and then predicting the masked region by sensing the speech context. It can solve  unnatural prosody in the edited region and synthesize the speech corresponding to the unseen words in the transcript. Secondly, for the possible operation of text-based speech editing, we design three text-based operations based on CampNet:  deletion, insertion, and replacement. These operations can cover various situations of speech editing. Thirdly, to synthesize the speech corresponding to long text in insertion and replacement operations, a word-level autoregressive generation method is proposed, which can synthesize the speech of arbitrary length text. Fourthly, we propose a speaker adaptation method using only one sentence for CampNet and explore the ability of few-shot learning based on CampNet, which provides a new idea for speech forgery tasks. The subjective and objective experiments\footnote{Examples of generated speech can be found at \href{https://hairuo55.github.io/CampNet}{https://hairuo55.github.io/CampNet.}} on VCTK and LibriTTS datasets show that the speech editing results based on CampNet are better than TTS technology, manual editing, and VoCo method (the combination of TTS and VC).  We also conduct detailed ablation experiments to explore the effect of the CampNet structure on its performance. Finally, the experiment shows that speaker adaptation with only one sentence can further improve the naturalness of speech editing for one-shot learning.

\end{abstract}


%
\IEEEpeerreviewmaketitle

\section{Introduction}
%
%
%
%
\IEEEPARstart{T}{}he rapid development of  internet has accelerated the transmission of information. There are various media for us to learn, entertain and communicate: movies, podcasts, YouTube videos, interactive online education, etc. The production of these media is often inseparable from speech editing. Typical speech editing interfaces \cite{derry2012pc} present a visualization of the speech such as waveform and/or spectrogram and provide the user with standard select, cut, copy, paste, and volume adjustment, which are applied to the waveform itself. Some advanced operations such as time-stretching, pitch bending, and de-noising are also supported. Such tools provide a great convenience for media producers and have a wide range of application scenarios \cite{whittaker2004semantic}.

\begin{figure}[tp]
    \centering 
    \includegraphics[width=8cm]{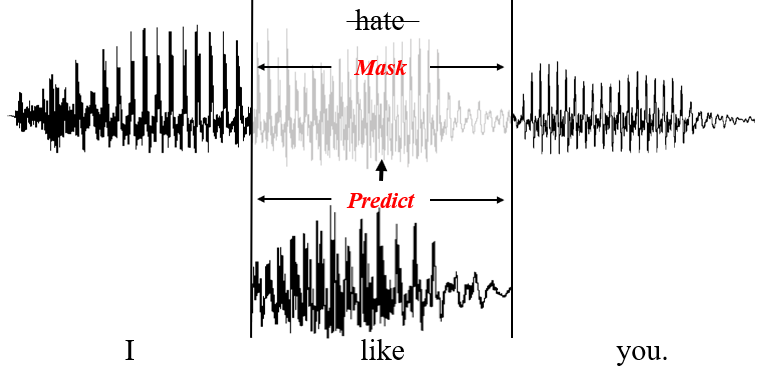}
    \caption{The replacement operation of text-based speech editing. The operation can be divided into two steps: first, masking the region to be edited and then predicting new speech according to the modified text and speech context. Deletion and insertion operations in text-based speech editing can also be divided into masking and prediction processes, as described in Sec. \ref{sec:operation}.}
    \label{fig:intro}
\vspace{-0.3cm}
\end{figure}

Some state-of-the-art systems allow the editor to perform select, cut, and paste operations in the text transcript of the speech and apply the changes to the waveform accordingly, which is called text-based speech editing \cite{jin2018speech}.  In many cases, it is useful to delete some words, insert a new word or phrase in the speech editing process, such as deleting dirty  words, replacing a misspoken word or inserting an adjective as emphasis. Text based speech editing can directly modify the speech waveform by deleting, inserting or replacing the target word in the text. However, it mainly faces two problems.  One is that the edited speech often sounds unnatural because the edited region does not match the prosody of speech context.
(e.g., mismatches in intonation, stress, or rhythm) \cite{morrison2021context}. Another is that the interfaces do not support the ability to synthesize new words not appearing in the transcript \cite{jin2018speech}. There are a series of studies on these problems.

For the first problem of prosody mismatch, the main reason is that the prosody of edited speech does not match the context of the original speech. Due to the continuity of human speech, there will be different prosody in different scenes. Directly concatenating the speech in different scenes will lead to problems such as discontinuity of fundamental frequency (F0) and background mismatch. To solve this problem, speech manipulation includes fast pitch-shifting and time-stretching techniques is applied, which include Time-Domain Pitch-Synchronous Overlap-and Add (TD-PSOLA) \cite{moulines1990pitch}, WORLD \cite{morise2016world}, and STRAIGHT \cite{kawahara2006straight}. These methods are efficient and suitable for real-time interactive applications but will produce audible artifacts \cite{airaksinen2018comparison,hu2013experimental}. Neural vocoders such as WaveNet \cite{oord2016wavenet}, etc. \cite{kalchbrenner2018efficient,prenger2019waveglow,yamamoto2020parallel,kumar2019melgan,kong2020hifi,valin2019lpcnet} can obtain higher perceptual quality than traditional methods, but can not perform context-aware generation for text-based speech editing. To generate prosodies that sound natural in context with any preceding or following speech, context-aware prosody correction \cite{morrison2021context} is applied to modify the prosodic information of the target segment. In this method, the prosodic information is predicted by neural network, then prosodic
modification is realized by applying the TD-PSOLA algorithm \cite{moulines1990pitch}, followed by de-noising and de-reverberation \cite{su2020hifigan}.
 This method combines neural network and digital signal processing method, which can effectively improve the speech quality. However, an obvious limitation of this system is that the words to insert or replace may not be found in the available speech data of the target speaker, which limits the application in the field of text-based speech editing.

 The second problem is that new words that do not appear in transcripts could not be synthesized.
It is easy for a person to type a new word not appearing in the transcript, but it is not obvious how to synthesize the corresponding speech. Of course, it is possible to record new audio with missing words, but it needs to access the original voice talent \cite{jin2018speech}, which will bring great difficulties to the speech editing process. With the rapid development of deep learning in the task of speech generation, the speech synthesized by machines can be comparable to humans, such as the works Tacotron \cite{wang2017tacotron,shen2018natural} and WaveNet \cite{oord2016wavenet} in the field of TTS.
Besides, some transfer learning works based on TTS can generate the speech of the target speaker, such as global style token (GST) \cite{wang2018style}, etc \cite{jia2018transfer, nachmani2018fitting,zhang2016end}. However, these methods are sentence level generation, and it is impossible to edit the specific words in the synthesis speech. To achieve text-based speech editing, the previous work was completed with the help of TTS and voice conversion (VC)  system \cite{9262021,mohammadi2017overview}, which is called VoCo \cite{jin2017voco,jin2018speech}. The key idea of VoCo is to synthesize the inserted word using a similar TTS voice (e.g., having correct gender) and then modify it to match the target speaker using the VC model  \cite{sun2016phonetic}(making an utterance by one person sound as if it was made by another). 
Since each module is independent of the other, it will accumulate errors and bring difficulties to the construction of the system.  
  EditSpeech \cite{tan2021editspeech} is developed upon a neural TTS framework. It uses force alignment technology to align text with speech first, and then uses partial inference and bidirectional fusion to incorporate the contextual information related to the edited region.  This method does not need pipeline structure,  but introduces a priori alignment information to realize the  mapping between edited text and target region, so as to avoid the unnatural phenomenon caused by cut-copy-paste operation.

Based on the analysis of the above two problems, it can be found that the operation of cut-copy-paste leads to the unnatural phenomenon of speech editing results. There are two reasons. First, it isn't easy to synthesize the edited region in combination with the speech context. Second, during the process of deleting, cutting or pasting some voice clips, it is easy to occur an unnatural connection.
Different from the cut-copy-paste operation, we view text-based speech editing as two steps. We take the replacement operation as an example, as shown in Fig. \ref{fig:intro}. The replacement process can be divided into two steps. First, we mask part of the original speech that needs to be edited. Then the masked region is predicted according to the modified transcription and speech context. In fact, some other operations of text-based speech editing, such as deletion and insertion, can also be viewed as the process of masking and prediction, which will be described in detail in Sec. \ref{sec:operation}. The advantage of this view is that text-based speech editing can be described by an end-to-end model, which can directly generate edited speech according to the context and ensure the natural prosody of speech.

This paper describes approaches to text-based speech editing that can automatically delete, replace and insert the speech at word level by editing the transcription. The approaches can avoid the unnatural problems caused by cutting and pasting in traditional methods and synthesize speech matching with the context in an end-to-end form. Firstly, the context-aware mask prediction network (CampNet) is proposed to simulate the process of text-based speech editing. Secondly, three text-based speech editing operations based on CampNet are designed: deletion, replacement, and insertion. Thirdly, a word-level autoregressive generation method is proposed to improve the editing length. 
Fourthly, a transfer learning method using only one sentence is proposed, which can further improve the naturalness of the model and provide a new idea for speech forgery.

Overall, the main contributions of this paper are:
 \begin{itemize}
\item We propose CampNet for the text-based speech editing task, which avoids the unnatural phenomenon caused by ``cut-copy-paste'' operation in the traditional method and can synthesize new words not appearing in the transcript.  To our best knowledge, CampNet is the first text-based speech editing model that can be trained in end-to-end form without duration information. The model uses mask prediction to simulate the text-based speech editing process (Sec. \ref{sec:mask}), and uses coarse-to-fine decoder to improve the perception of speech context (Sec. \ref{sec:decoder}).
\item Based on CampNet, we design three speech editing operations, corresponding to delete, replace and insert operations, respectively (Sec. \ref{sec:operation}). These operations can comprehensively cover different kinds of situations that text-based speech editing can face.
\item In the text-based insertion and replacement operations, to synthesize the speech corresponding to long text, a word-level autoregressive generation method is proposed (Sec. \ref{sec:word-level}), which can synthesize the speech of arbitrary length text by using CampNet.
\item We propose a one-sentence speaker adaptation method for the CampNet, further boosting the performance for one-shot learning, and providing a new idea for speech forgery. Experiments show that this method can obtain better speech similarity for unseen speakers with only one sentence than TTS and VC systems (Sec. \ref{sec:shot}).
\end{itemize}

\begin{figure*}[htp]
    \centering 
    \includegraphics[width=18cm]{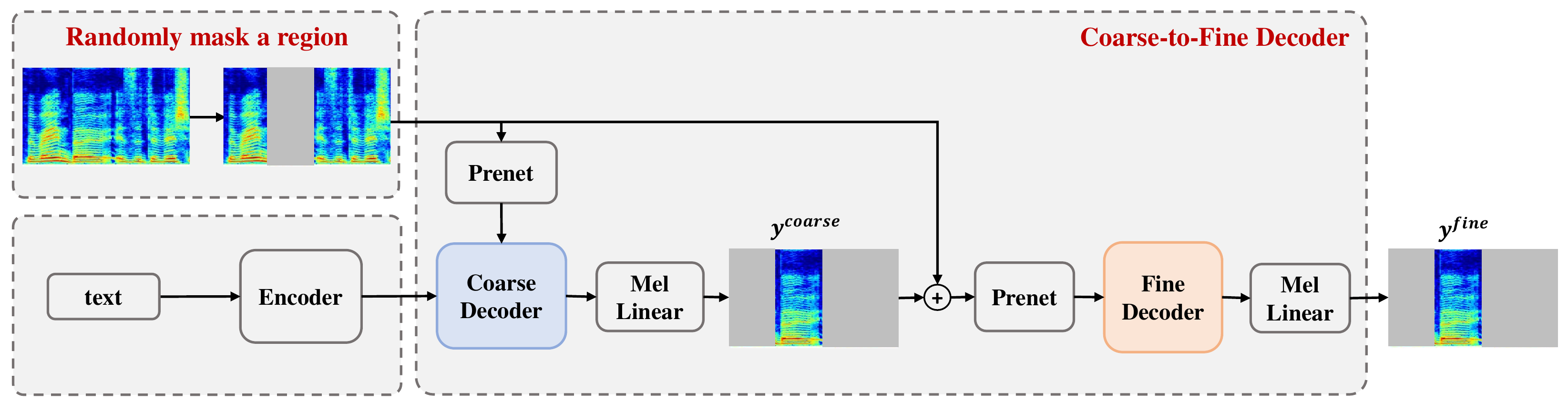}
    \caption{Structure of \textbf{C}ontext-\textbf{A}ware \textbf{M}ask \textbf{P}rediction \textbf{Net}work (CampNet). To simulate text-based speech editing, the model randomly masks a region of speech at the training stage and then predicts the masked speech according to the corresponding text and speech context.}
    \label{fig:campenet}
\vspace{-0.3cm}
\end{figure*}

This paper is structured as follows. The proposed CampNet, its operations for text-based speech editing, word-level autoregressive generation method, and its transfer learning method are described in Section \ref{sec:2}. After explaining the experiments and results in Section \ref{sec:3} and Section \ref{sec:4}, we draw a conclusion in Section \ref{sec:5}.
\vspace{-0.1cm}

\section{Context-Aware Mask Prediction Network}\label{sec:2}

The context-aware mask prediction network (CampNet) consists of two processing stages, as shown in Fig. \ref{fig:campenet}: encoder and decoder. First, the encoder module processes the input sentence and converts it into a hidden representation. This representation is used to guide the decoder to predict the acoustic feature of the edited speech. Second, a random region of acoustic features is masked as the ground truth to condition the decoder at the decoding stage. The decoder is divided into two steps. The first step is to learn the alignment information between the masked ground truth and the text representation through the multi-head attention mechanism and predict coarse acoustic features. Then, the second step is to predict finer acoustic features based on the coarse acoustic features and original speech context, which can further fuse the context information of speech to make the predicted speech more natural. We call the process of masking part of the acoustic features and predicting the mask region as the "context-aware mask prediction".

In the section, we will first introduce CampNet. Secondly, we will show how to use CampNet for text-based speech editing tasks, including delete, replace and insert operations. Thirdly, the word-level autoregressive generation method is proposed. Finally, we will present the few-shot and one-shot learning methods based on CampNet.

\subsection{Context-Aware Mask Prediction}\label{sec:mask}
The task  of end-to-end text-based speech editing model is to modify part of the original speech to match the edited transcription. Given the source acoustic features $y = (y_1, \dots, y_n , \dots, y_m, \dots,  y_T)$  and its transcription sequences $x = (x_1,\dots, x_a, \dots,  x_b,  \dots,x_M)$, where  $(x_a, \dots,  x_b)$ is aligned with  acoustic features $(y_n , \dots, y_m)$.  When $(x_a, \cdots,  x_b)$ in transcription $x$ is edited and the new transcription $x'$ is $(x_1,\dots, x'_a, \dots,  x'_{b'},  \dots,x_{M})$,
the target acoustic feature is $y' = (y_1, \dots, y'_{n} , \dots, y'_{m}, \dots,  y_T)$.
We assume that the length of the new speech in the editing region is consistent with that of the original region in the training stage. If it is inconsistent at the inference stage, an additional duration model \cite{wu2016merlin} can be used to predict the duration of edited words. Then we can add or delete some fragments on the original part to achieve consistent length. The purpose of this assumption is to ensure that there is no mismatch between the model in the training stage and the test stage.
So, the problem of text-based speech editing can be formulated in terms of estimating the condition probability $P(y'|y,x,x';\theta)$ of the target acoustic feature, and $\theta$ is corresponding model parameters, where
\begin{equation}
P(y'|y,x,x';\theta) =  P(y'_{n} , \dots, y'_{m}|y,x,x';\theta)\label{eq1}
\end{equation}

\begin{figure*}[tp]
    \centering 
    \includegraphics[width=17.5cm]{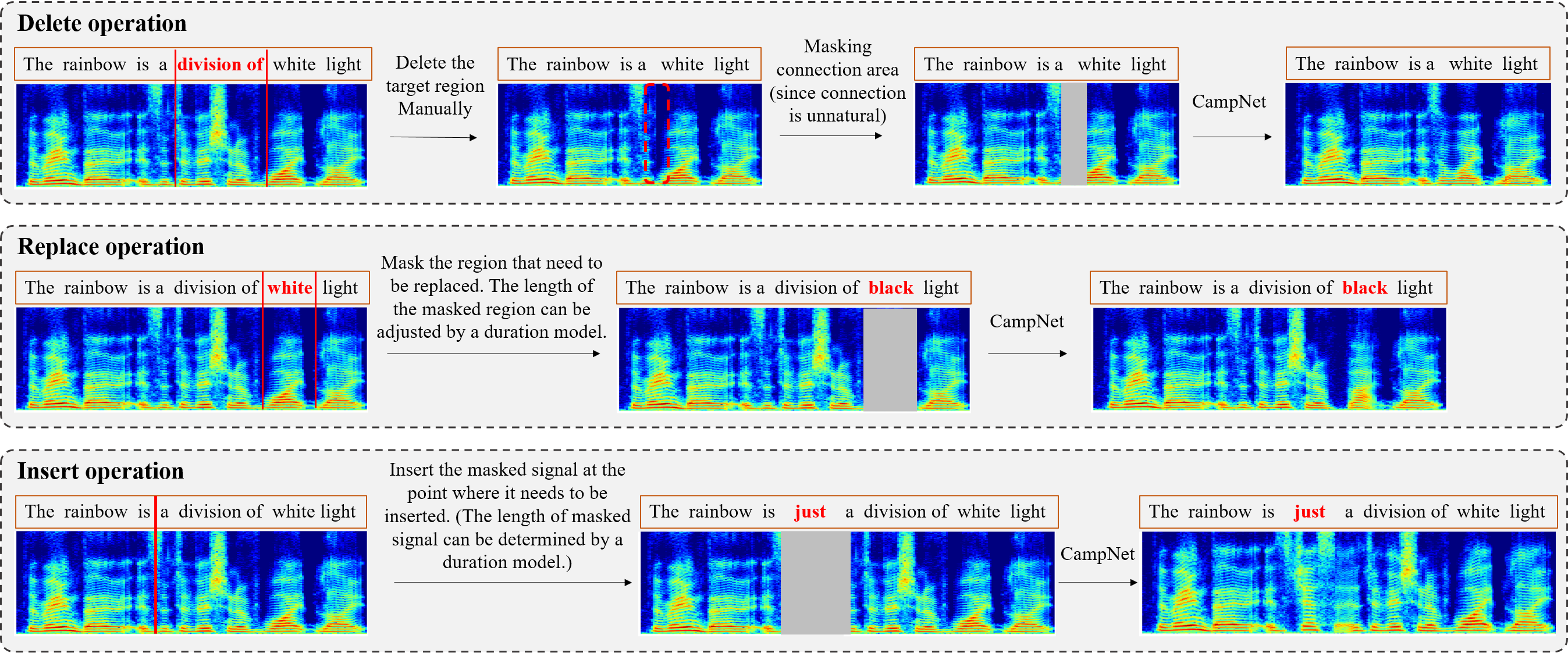}
    \caption{At the inference stage, three different operations of text-based speech editing based on CampNet are proposed.}
    \label{fig:operation}
\vspace{-0.3cm}
\end{figure*}

Since $(y'_{n} , \dots, y'_{m})$ is independent of text  sequences $x$ and  related to text $x'$, we can get the formula from Eq. \ref{eq1}:
\begin{equation}
P(y'|y,x,x';\theta) =  P(y'_{n} , \dots, y'_{m}|y,x';\theta)\label{eq2}
\end{equation}

In addition, it can be observed that  $(y'_{n} , \dots, y'_{m})$ and $(y_{n} , \dots, y_{m})$  have the same position, and their contents are different. To remain the position information of edited region $(y'_{n}, \dots, y'_{m})$ and  not be interfered by the content information of $(y_{n} , \dots, y_{m})$, $(y_{n} , \dots, y_{m})$ can be replaced with a new token $\mathrm{<mask>}$, and the masked source acoustic features $y_\mathrm{mask}$ can be expressed as:
\begin{equation}
y_{\mathrm{mask}} = (y_1, \dots, \mathrm{<mask>} , \dots, \mathrm{<mask>}, \dots,  y_T)
\end{equation}
Where the $n$ position in $y_{\mathrm{mask}}$ is the starting point of the mask, and $m$ is the ending point of the mask.

Then the  condition probability $P(y'|y,x,x';\theta)$ can be formulated as:
\begin{equation}
P(y'|y,x,x';\theta) =  P(y'_{n} , \dots, y'_{m}|y_{\mathrm{mask}},x';\theta)\label{eq3}
\end{equation}

From the Eq. \ref{eq3}, the task of text-based speech editing can be decomposed into the following two process. First, mask the region of the original  speech $y$ that needs to be edited, and  get the masked acoustic features $y_\mathrm{mask}$. Then, combined with the masked acoustic feature $y_\mathrm{mask}$ and  the edited text sequence $x'$, neural network is used to predict the edited region  $(y'_{n} , \dots, y'_{m})$.
Because $x'$ and $y_\mathrm{mask}$ have different lengths, and it has been proved that transformer can effectively fuse the context information of sequences with different lengths \cite{li2019neural}, an encoder-decoder framework based on transformer is adopt as the structure of CampNet, which is shown in Fig. \ref{fig:campenet}.

In the CampNet, an encoder processes the input sentence $ x' =  (x_1,\dots, x'_{a}, \dots,  x'_{b'},  \dots,x_M)$ and converts it into a hidden representation in the following way:
\begin{equation}
m=\left(m_{1}, m_{2}, \ldots, m_{M}\right)=\operatorname{encoder}_{\theta_{e}}(x')\label{eq4}
\end{equation}
where $\theta_{e}$ denotes the parameters of encoder network. 

Since there are only text and speech pairs in the training stage, we randomly mask part of the speech and then use the network to predict the masked part to simulate the text-based speech editing process. The advantage of the masking mechanism is that the training data of speech editing can be well simulated without parallel corpus of speech editing. In addition, due to randomly mask a region during training, a wealth of augmented data can be obtained, improving the robustness of the model.
The masked acoustic features $y_\mathrm{mask}$ are first consumed by a neural network composed of two fully connected layers with ReLU activation \cite{nair2010rectified}, named $prenet$. It is responsible for projecting $y_{\mathrm{mask}}$ into the same subspace as phoneme embeddings, which can be expressed as:
\begin{equation}
h=\left(h_{1}, h_{2}, \ldots, h_{T}\right)=\operatorname{prenet}_{\theta_{p}}(y_{\mathrm{mask}})\label{eq5}
\end{equation}
where $\theta_{p}$ denotes the parameters of $prenet$.

Finally, combined with $m$ and $h$ hidden features, the final signal is predicted by the decoder, which can be expressed as:
\begin{equation}
\left(y_{n'}, \ldots, y_{m'}\right)=\operatorname{decoder}_{\theta_{d}}(h,m)\label{eq5}
\end{equation}
where $\theta_{d}$ denotes the parameters of decoder network.

At the decoding stage, how to perceive the context of speech is the key to synthesize natural speech. We propose a coarse-to-fine decoding method to achieve context-aware, and the details are introduced in the following subsection.

\subsection{Coarse-to-Fine Decoding} \label{sec:decoder}
\begin{figure}[tp]
    \centering 
    \includegraphics[width=7cm]{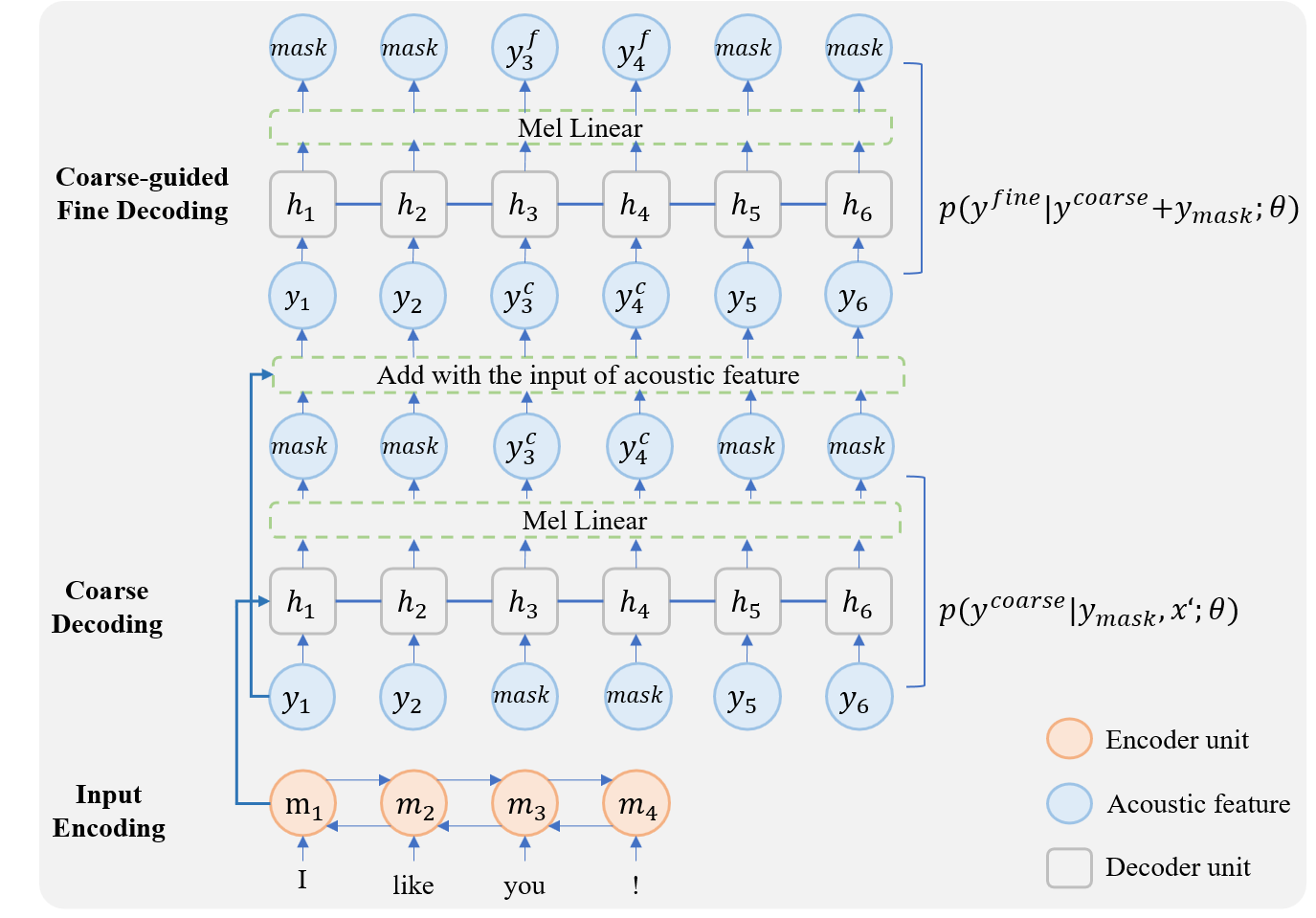}
    \caption{Structure of coarse-to-fine decoding. We first generate the coarse
acoustic features $y^{\mathrm{coarse}}$ from text information and masked acoustic features. Then, a fine decoder fills in the missing details based on coarse acoustic features and speech context, and predicts finer acoustic feature $y^{\mathrm{fine}}$.}
    \label{fig:decoder}
\end{figure}

To better perceive context information of speech and make the predicted speech more natural, we propose a two-stage decoder, which is named as “coarse-to-fine decoding”. The structure is a two-stage transformer in series, as shown in Fig. \ref{fig:decoder}. Since our framework is non-autoregressive, when we perform the first decoding (coarse decoding), the model needs to predict all frames of the mask area at one time, and can not fuse the information of the previous frame in the form of autoregression to predict the information of the next frame. Therefore, to better integrate the context information, we perform the second decoding (fine decoding) based on the accumulation of the output of the first-level decoder and the original masked speech information. Then the model can obtain better context information, rather than directly predict the speech information from the masked signal.

Since the transformer can output the whole sequence in parallel, 
the coarse decoding process can be represented by the product of the probabilities of each frame:
\begin{equation}
P(y^{\mathrm{coarse}} |y_{\mathrm{mask}},x';\theta) = \prod_{t=1}^{T} P\left(y_{t} \mid y_{\mathrm{mask}},x' ; \theta\right) \label{eq:coarse}
\end{equation}
In this process, we only predict the acoustic features of the masked region, while the output target of the region without masking is the mask value $\mathrm{<mask>}$, as shown in Fig. \ref{fig:decoder}. 
It can be seen from Eq. \ref{eq:coarse} that each frame of $y^\mathrm{coarse}$ is output in parallel. Compared with the autoregressive models such as Tacotron and WaveNet, the non-autoregressive model can not fuse the information of the previous frame well. This will cause the naturalness of the synthesized speech to be worse than that of the autoregressive model.
To solve this problem, we perform the second decoding, which feeds the sum of $y^{\mathrm {coarse }}$ and $y_{\mathrm {mask }}$ to a fine decoder to predict finer acoustic features $y^{\mathrm {fine}}$. It can be expressed as:
\begin{equation}
P\left(y^{\mathrm{fine }} \mid y^{\mathrm {coarse }}, y_{\text {mask }} ; \theta\right)=\prod_{t=1}^{T} P\left(y_{t} \mid y^{\mathrm {coarse }}+y_{\mathrm {mask }} ; \theta\right) 
\end{equation}

The second decoding can combine the output of the first-level decoder and the context information of original speech, which is helpful to generate more natural and expressive speech. The target of the second decoder is the same as the first decoder. The advantage of this is that the model can synthesize speech in a non-autoregressive way, and better integrate context information in the form of two-level decoding.


\subsection{Speech Editing Operations Based on CampNet}\label{sec:operation}
With a pre-trained CampNet model, some operations of speech editing, such as deletion, insertion, and replacement, can be carried out. The operations are shown in Fig. \ref{fig:operation}. In this section, we will introduce these in detail.
\subsubsection{Delete operation}
The deletion operation allows the user to remove a region of speech waveform that corresponds to certain specified words. We divide the process into three steps. The first step is to manually delete the target region and the corresponding words in the text. 
Due to manual deletion, unnatural phenomena will appear at the connection, such as fundamental frequency discontinuity. To repair this problem, there are two solutions. One is convenient but empirical, and the other is accurate but requires additional duration model. For the former, we can take the connection point as the center and mask the left and right of speech in a small range. Then, we input the masked speech and the text after deleting the target word into CampNet to re-predict the masked region. The reason for masking a small range at the connection is to re-predict the pronunciation of word-final of the previous word and word-initial of the next word. Since masking a small area at the connection is empirical, another more accurate method is to use an additional duration model to predict the pronunciation duration of word-final of the previous word, word-initial of the next word, as well as the pause information of the two words. Then mask a specific range according to this information. Users can adopt one of two methods according to the needs of different scenarios.
\subsubsection{Replace operation}
The replace operation allows the user to replace a fragment of speech with another speech. The operation can be divided into two cases. One is that the length of the replaced segment is close to the target pronunciation. The other is that there is a large gap between them. For the former, it can be divided into two steps. The first step is to define the word boundary to be replaced, mask it according to the word boundary and then modify the text. It is worth noting that the range of masking can be larger than the actual boundary when masking. In this way, the model can learn more natural connections. The second step is to input the masked speech and the modified text into CampNet. The model will predict the replaced speech according to the modified text.

If there is a big difference between the length of the replaced speech and the original speech, such as adding some words or deleting some words, a pre-training duration model can be used to predict the length of the replaced region. The duration model is widely used in traditional TTS  task\cite{wu2016merlin}. Here, we use the duration model to obtain the speech length of the replaced word.  Then according to the predicted length, the masked region can be added or deleted some fragment to ensure the consistency of the duration.

\subsubsection{Insert operation}
The insert operation allows the user to insert a speech into the edited speech. This operation is similar to the replacement operation. Firstly, we can use the pre-trained duration model to predict the duration of the words to be inserted. Then insert the masked signal with the predicted length into the original speech. Finally, input the modified text and speech into CampNet, and predict the inserted speech.

It is worth mentioning that when we insert or replace some words, to make the pronunciation of the edited words more natural with adjacent words, we can mask part of the pronunciation of the adjacent words appropriately, such as the word-final of the previous word, and word-initial of the next word. Then, CampNet is used to re-predict the pronunciation of the masked area of these adjacent words, so as to make the prosodic connection more natural.

\begin{figure}[tp]
    \centering 
    \includegraphics[width=6cm]{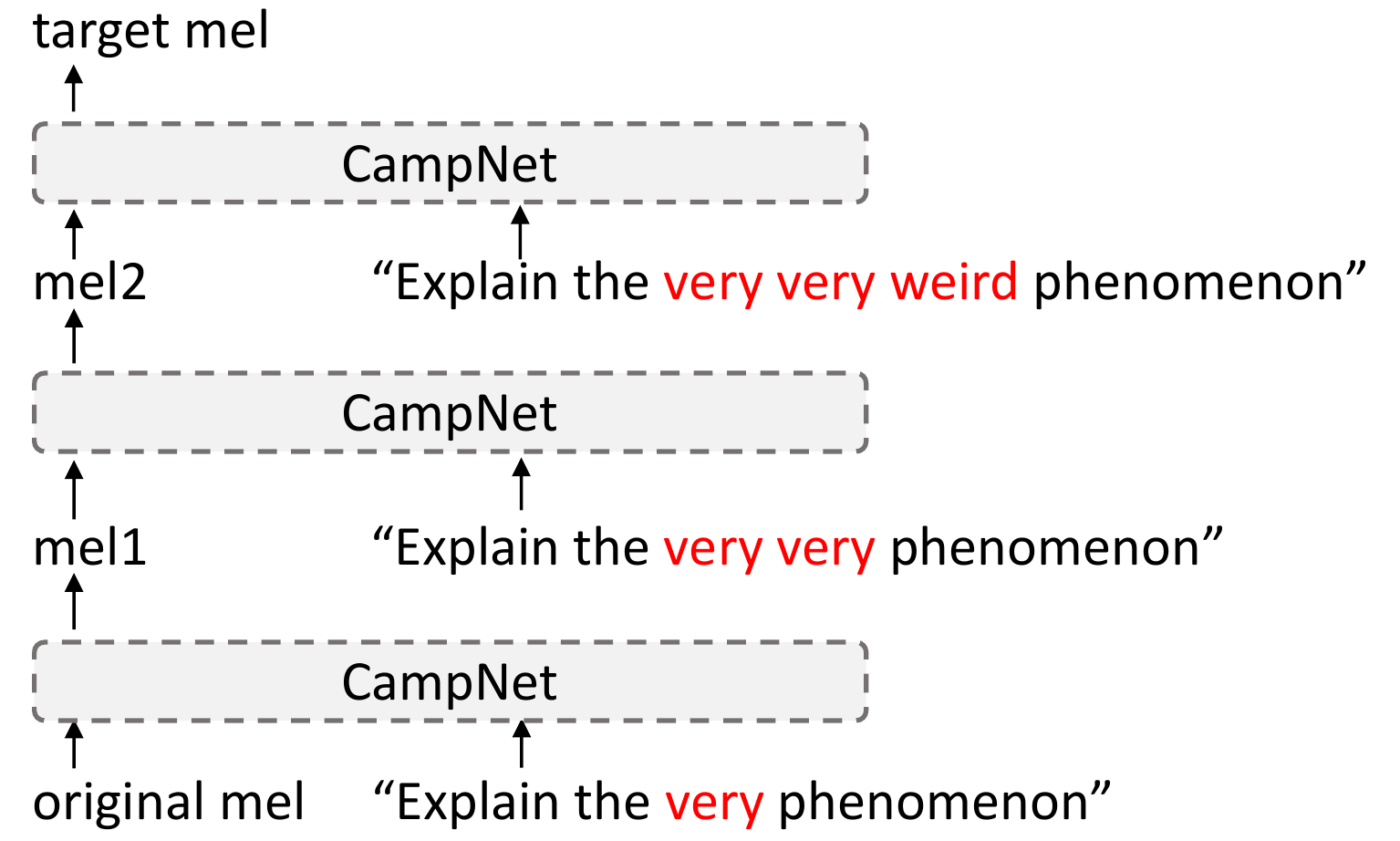}
    \caption{An example of word-level autoregressive generation method. The figure shows the process of inserting a speech with the text "very very weird" (marked in red) into the original speech.}
    \label{fig:wordlevel}
\vspace{-0.5cm}
\end{figure}

\subsection{Word-level Autoregression Generation Based on CampNet}\label{sec:word-level}
In the replacement or insertion operation of text-based speech editing in Sec. \ref{sec:operation}
, it is easy to face the need to  generate speech  with long text (For example, the text to be replaced has many words, or the inserted speech has many words).  Since we only mask a small part of speech in the training process, when we need to generate speech with long text in the inference stage, the training and inference stages do not match, and the performance of the model will be poor.  Therefore, in order to enable CampNet to synthesize long text speech in the inference stage, we propose a word-level generation method, which can effectively solve this problem.

The generation method based on word-level autoregression is shown in the Fig. \ref{fig:wordlevel}, different from synthesizing the speech corresponding to all texts at one step, we view the generating long text speech as a multi-stage process. First, insert the first word into the original text and synthesize the speech with the modified text. Then, on this basis, insert the second word and cycle in turn until all words are inserted, then the final speech can be obtained. Since this process generates speech word by word, it is called word-level autoregressive generation method.

It should be noted that the word-level autoregressive generation is important for CampNet. This method can make CampNet not limited to the problem of editing length, but can synthesize speech of any length and ensure the stability of generation.

\begin{figure}[tp]
    \centering 
    \includegraphics[width=6.5cm]{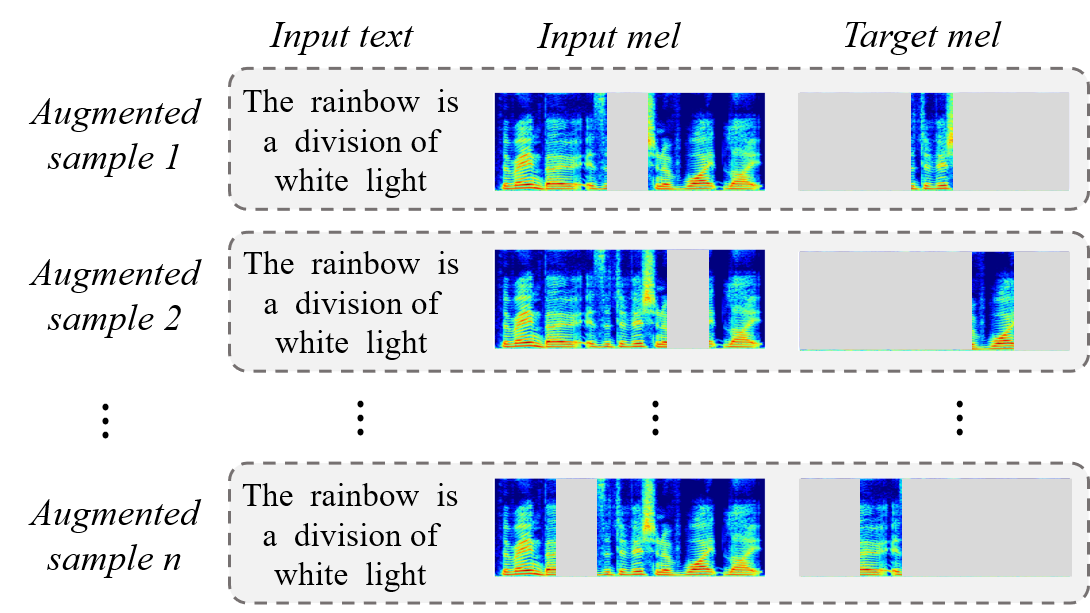}
    \caption{An example of data augmentation using only one utterance. By randomly masking a region of speech, one sample can be extended to multiple different inputs.}
    \label{fig:oneshot}
\vspace{-0.5cm}
\end{figure}
 \subsection{Transfer Learning for Text-based Speech Editing}\label{sec:shot}
When the target speaker's corpus is small, synthesizing the target speaker's voice has always been a research hotspot. Many studies are conducted from the perspective of TTS and VC. Unlike TTS and VC systems, the proposed framework uses the existing speech as input to modify individual words. Compared with synthesizing a whole sentence simultaneously, the task is simpler, so it is easy to get better stability and similarity. In addition, CampNet can obtain the speaker information through the unmasked region without using the speaker embedding, which avoids the error caused by the extraction of the speaker embedding.

After pre-training the CampNet with multi-speaker data, the model can synthesize a new speaker's voice without transfer learning. We can also fine-tune the  parameters for better results, similar to the transfer learning in TTS and VC tasks. However, CampNet can enlarge the training data by randomly masking different positions of speech. Even the training corpus has only one sentence, which can not be achieved by the TTS and VC model. This section will introduce the transfer learning methods based on CampNet for few-shot learning and one-shot learning in detail.

\subsubsection{Few-shot learning}\label{sec:few-shot}

When we need to use a small corpus for speaker adaptation, we only need to transfer the parameters related to speaker features in our proposed framework. Since the encoder only has text input and there is no speaker embedding to control the speech, after pre-training the model with a multi-speaker dataset, we can consider the encoder's parameters universal. We only need to fine-tune the decoder because the input of the decoder is the acoustic features. After fine tuning the decoder, the model can predict the acoustic features that better match the target speaker.

\subsubsection{One-shot learning}\label{sec:one-shot}
One-shot learning has always been the difficulty of speech forgery. This section provides a different idea from TTS and VC, which can obtain better similarity. We can directly fine-tune the model with only one sentence to improve the similarity with the target speaker who has never appeared in the training corpus. Because CampNet is based on mask and prediction mechanism, randomly masking acoustic features can effectively augment the training data and improve the robustness of the model, which is shown in Fig. \ref{fig:oneshot}. After several steps of fine-tuning the model with one sentence, the model is not easy to overfit, and the performance will be further improved due to the difference of input at each step.

\vspace{-0.1cm}
\section{EXPERIMENTAL PROCEDURES}\label{sec:3}
\vspace{-0.1cm}
\subsection{Dataset and Task}
In this section, we conduct experiments on VCTK \cite{veaux2017cstr} and LibriTTS \cite{zen2019libritts} corpus to evaluate our proposed method. The VCTK corpus includes speech data uttered by 110 English speakers with different accents. Each speaker reads out about 400 sentences. Specifically, we select four speakers from the VCTK dataset as the test set, and the rest utterances are divided into 90\% training set and 10\% validation set. We use the training set to train the model. We also randomly select 100 sentences from the LibriTTS corpus as the test set to verify the model's performance on the cross dataset.

We mainly compare the replacement operation of CampNet and other systems, which is easy to evaluate with the original speech \footnote{Examples of more operations can be found at \href{https://hairuo55.github.io/CampNet}{https://hairuo55.github.io/CampNet.}}. To ensure that the edited speech of different systems is consistent with the content of the original speech, which facilitates the comparison between objective metrics and subjective metrics, we randomly choose 80 words that span 3 to 10 phonemes from 80 different sentences for each  test set. For each sentence, we remove the region of the corresponding words in the speech. Then we use different systems to predict the removed region. All wav files are sampled at 16KHz.
\vspace{-0.3cm}
\subsection{Model Details}
\begin{figure}[tp]
    \centering 
    \includegraphics[width=8cm]{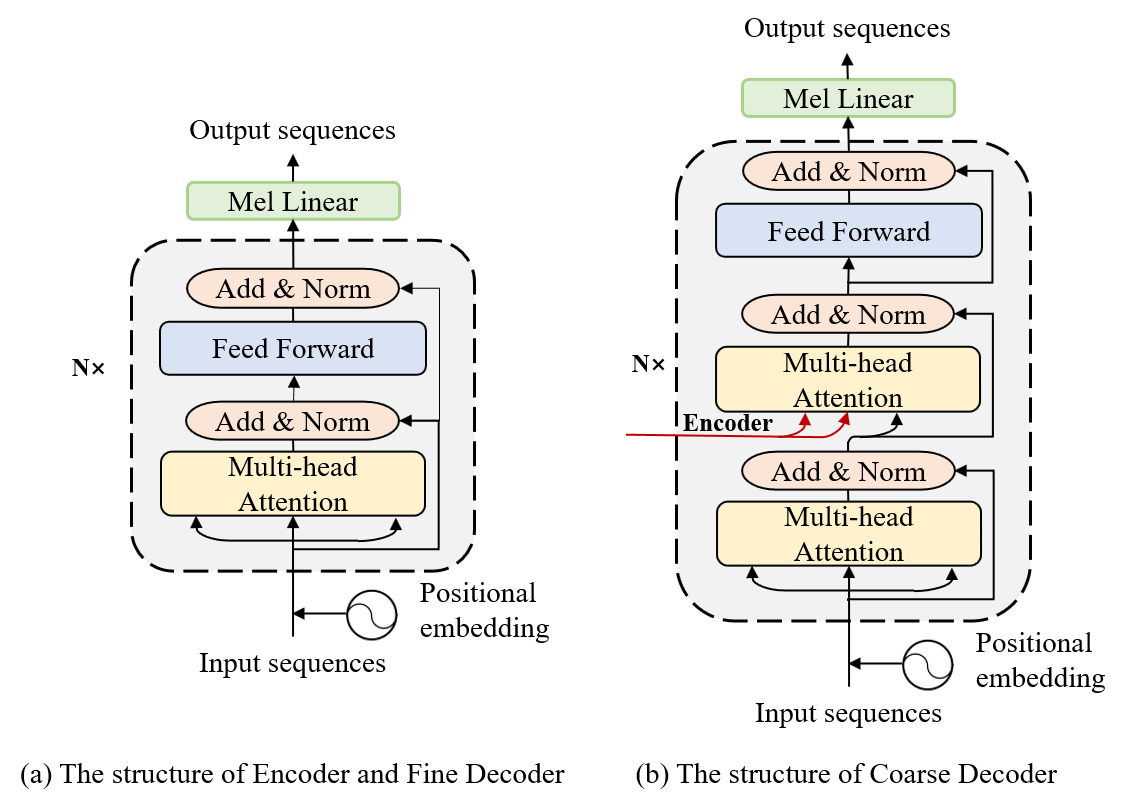}
    \caption{Structures of encoder, coarse decoder and fine decoder of CampNet.}
    \label{fig:transformer}
\vspace{-0.4cm}
\end{figure}

Acoustic features are extracted with a 10 ms window shift. LPCNet \cite{valin2019lpcnet} is utilized to extract 32-dimensional acoustic features, including 30-dimensional BFCCs \cite{gulzar2014comparative}, 1-dimensional pitch and 1-dimensional pitch correction parameter. Five systems are compared in our experiments, including TTS technology,  the combination of TTS and VC, manual editing, CampNet and actual recording.  We use the training set in VCTK to train the LPCNet model. 
\begin{itemize}
  \item \textbf{Synth} We train a neural TTS system to synthesize the speech and copy the target region to insert into the edited speech. To make the voice of the synthesized speech as similar as that of the target speaker as possible, Tacotron2  based on global style token (GST \cite{wang2018style}) is used as the acoustic model. The structure of Tacotron2 is the same as that in  paper \cite{shen2018natural}. The hidden dimension of the  encoder and decoder in Tacotron is  512, and the GST dimension is  128.
We use the phoneme as input and output the 32-dims acoustic features extracted by LPCNet. The initial learning rate is set to 1e-3. Adam \cite{kingma2014adam} is used as the optimizer.
  \item  \textbf{VoCo}  The main idea of VoCo is to synthesize the inserted word using a similar TTS voice (e.g., having correct gender) and then modify it to match the target voice using a VC model. To realize the VoCo system, we train the neural TTS system and VC system, respectively. For the TTS system, the configuration is the same as \textbf{Synth}. We select two speakers (one male and one female) from the training set as source speakers. Then, the TTS system is used to synthesize speech, and the VC system synthesizes the voice which is similar to the target speaker. We copy the target region from the output speech and insert it into the edited speech. The voice conversion is based on phonetic posteriorgrams (PPGs) \cite{sun2016phonetic}, which can be applied to non-parallel voice conversion and achieved both high naturalness and high speaker similarity of the converted speech. The 512-dimensional PPGs are extracted from the acoustic model in speaker independent-ASR, which is implemented using the Kaldi toolkit \cite{povey2011kaldi} and trained on our 20,000 hours corpus. The voice conversion model’s structure follows the structure in paper \cite{sun2016phonetic}. Furthermore, to make the voice conversion system have the one-shot ability, we use GST as speaker embedding to train a multi-speaker VC model.
The initial learning rate of the VC model is set to 1e-3, and Adam \cite{kingma2014adam} is used as the optimizer.  It should be noted that the TTS technology in the original VoCo adopts the unit selection technology, which is complex to realize in VCTK datasets. While, at present, the popular end-to-end TTS technology can also generate natural speech. Therefore, we mainly adopt the idea of realizing speech editing with TTS and VC technology in VoCo, and use Tacotron to realize TTS. 
  \item \textbf{Edit} We use the editing interface to refine the speech further if it improves on Synth/VoCo.
  \item \textbf{Real} The actual recording without modification.
  \item \textbf{CampNet} The proposed framework of CampNet is shown in Fig. \ref{fig:campenet}. The structures of encoder, coarse decoder and fine decoder are based on transformer \cite{vaswani2017attention}, as shown in Fig. \ref{fig:transformer}. We input the phoneme sequence into a 3-layer CNN \cite{goodfellow2016deep} to learn the context information of the text. Each phoneme has a trainable embedding of 256 dims, and the output of each convolution layer has 256 channels, followed by batch normalization and ReLU activation and a dropout layer as well \cite{ioffe2015batch,nair2010rectified,srivastava2014dropout}. The transformer blocks of the encoder and fine decoder are 3. The transformer block of the coarse decoder is 6. The hidden dimension of the transformer is 256. At the training stage, we set the masked region to be 12\% of the total speech length. The initial learning rate is 1e-3. Adam \cite{kingma2014adam} is used as the optimizer. All batches are set to 16, and the number of training steps is 2 million.
\vspace{-0.1cm}
\end{itemize}

\begin{table*}[t]
    \centering
        \caption{OBJECTIVE EVALUATION RESULTS OF  SYNTH, VOCO, EDIT AND
CAMPNET ON THE TEST SETS OF VCTK AND LIBRITTS}
\scalebox{1.0}{
\begin{tabular}{c|cccc|cccc}
\hline \hline    &  & \multicolumn{2}{c}{VCTK}   & & & \multicolumn{2}{c}{LibriTTS}    & \\ \hline
Metrics & \textbf{Synth}\ & \textbf{VoCo} & \textbf{Edit}  & \textbf{CampNet} & \textbf{Synth}\ & \textbf{VoCo} & \textbf{Edit}  & \textbf{CampNet}   \\
\hline   MCD(dB)  & 0.594 & 0.589   &  0.582 &\textbf{0.380}   &  0.871&  0.894   & 0.870 & \textbf{0.628 }\\
 F0-RMSE(Hz)    &   10.463  &   10.555 &10.451  & \textbf{8.637}  &21.898 & 2.093   &  21.308 &\textbf{20.201} \\
    V/UV error  &1.944 & 1.843     & 1.937 & \textbf{1.635}  &3.916& 4.347  &3.956 &\textbf{3.675}  \\ 
  F0-CORR  &   0.973 &0.972    &  0.975& \textbf{0.981}  &0.940&0.945  &0.939&\textbf{0.954}  \\
\hline \hline
\end{tabular}}
\label{table:a1}
\vspace{-0.3cm}
\end{table*}
\begin{figure*}[tp]
    \centering 
    \includegraphics[width=17cm]{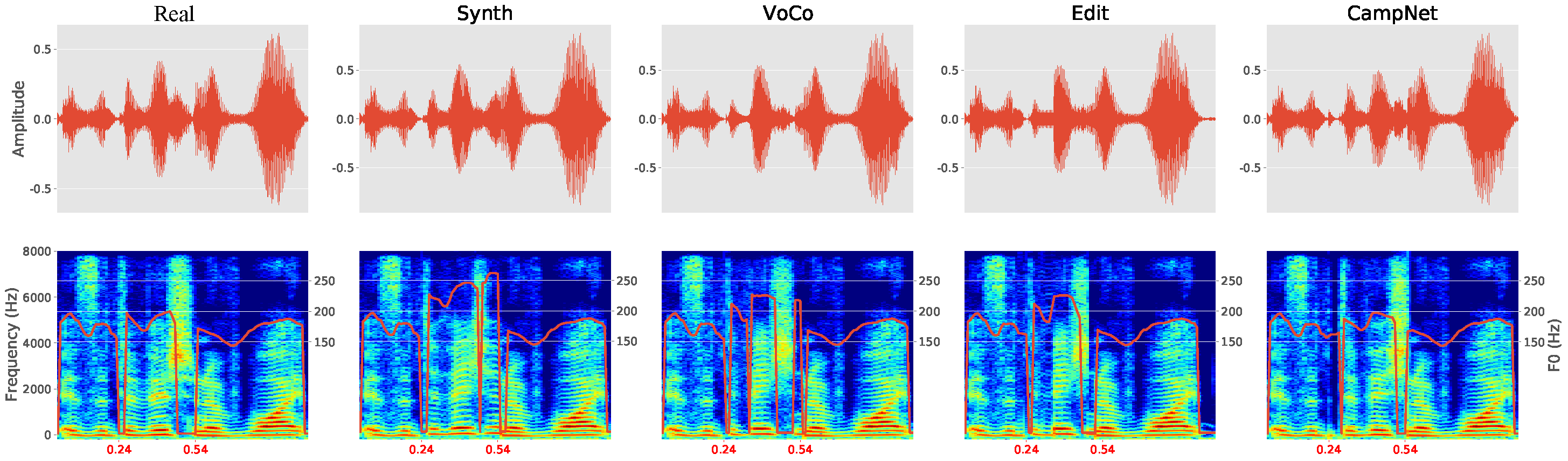}
    \caption{The waveforms and spectrograms of natural speech and speech edited by different system (the speaker did not appear in the training set). The region marked with time ($0.24s \sim 0.54s$) is the edited region. The text of the masked region is 'division'.}
    \label{fig:figure2}
\vspace{-0.4cm}
\end{figure*}

\vspace{-0.2cm}
\subsection{Objective Evaluation Metrics}
The quality of edited speech can be evaluated by comparing it with the actual speech using the following metrics. To reflect the overall information of speech (including the information of the editing area and the information of the junction), we calculate the objective metrics of the whole speech. Specifically, since CampNet only synthesizes acoustic features, speech waveforms need to be generated by vocoders. In contrast, other systems paste the target speech directly into the original speech waveform. Therefore, to avoid the influence of vocoder, we also paste the target region predicted by CampNet to the corresponding position of the original speech when calculating the objective metrics. Such objective comparison can be more accurate and fair. In this way, the speech of the unedited area of all systems can be guaranteed to be consistent. In addition, the test set of each system is the same, so it can be ensured that the metrics are only related to the speech of the editing area, and it will not cause unfairness in the comparative experiment. 

\subsubsection{Mel-Cepstral Distortion (MCD)}
Given two mel-cepstra $\hat{\mathbf{x}}=\left[\hat{x}_{1}, \ldots, \hat{x}_{M}\right]^{\top}$  and $ \mathbf{x}=\left[x_{1}, \ldots, x_{M}\right]^{\top}$, we use the mel-cepstral distortion (MCD) \cite{kubichek1993mel}:
\begin{equation}
\mathrm{MCD}[\mathrm{dB}]=\frac{10}{\ln 10} \sqrt{2 \sum_{i=1}^{M}\left(\hat{x}_{i}-x_{i}\right)^{2}}
\end{equation}
to measure their difference. Where  $M$ is the order of mel-cepstrum and $M$ is 28 in our  implementation.   Here, we used the average of the MCDs taken along the DTW \cite{muller2007dynamic} path between edited and reference feature sequences as the objective performance measure for each test utterance.
\subsubsection{F0-RMSE (Hz)}
For the F0 of speech, following RMSE is applied \cite{ai2020neural}:
\begin{equation}
\operatorname{F0-RMSE}=1200 \sqrt{(\left(\log _{2}\left(F_{r}\right)-\log _{2}\left(F_{s}\right)\right)^{2}}
\end{equation}
where the subscript $r$ and $s$ represent reference and edited speech, respectively.  The  F0-RMSE is calculated for each frame, and we use the average of the F0 taken along the DTW path  between converted and reference feature sequences.
\subsubsection{V/UV error}
The ratio of the number of unmatched U/V frames between reference and edited speech to total frames is calculated as the V/UV error \cite{ai2020neural}. For two different lengths of speech, we still use the DTW algorithm to align them.
\subsubsection{F0-CORR}
We use the correlation coefficient between the edited and reference F0 contours as the objective performance measure to evaluate the F0 of edited speech \cite{ai2020neural}. Since the synthesized and reference speech are not necessarily aligned in time, we computed the correlation coefficient after properly aligning them using the DTW algorithm. If we use $\tilde{\mathbf{y}}=\left[\tilde{y}_{1}, \ldots, \tilde{y}_{M^{\prime}}\right] $ and $ \mathbf{y}=\left[y_{1}, \ldots, y_{M^{\prime}}\right]$ to denote the vectors consisting of the elements which is aligned. We can use the correlation coefficient between $\tilde{\mathbf{y}}$ and $\mathbf{y}$
 \begin{equation}
R=\frac{\sum_{m^{\prime}=1}^{M^{\prime}}\left(\tilde{y}_{m^{\prime}}-\tilde{\varphi}\right)\left(y_{m^{\prime}}-\varphi\right)}{\sqrt{\sum_{m^{\prime}=1}^{M^{\prime}}\left(\tilde{y}_{m^{\prime}}-\tilde{\varphi}\right)^{2}} \sqrt{\sum_{m^{\prime}=1}^{M^{\prime}}\left(y_{m^{\prime}}-\varphi\right)^{2}}}
\end{equation}
where $\tilde{\varphi}=\frac{1}{M^{\prime}} \sum_{m^{\prime}=1}^{M^{\prime}} \tilde{y}_{m^{\prime}} $ and $ \varphi=\frac{1}{M^{\prime}} \sum_{m^{\prime}=1}^{M^{\prime}} y_{m^{\prime}}$, to measure the similarity between the two F0 contours.


\section{Results}\label{sec:4}
In this section, we first compare the performance of CampNet and some other systems, such as objective metrics, subjective metrics, and operating efficiency. Then, some ablation experiments based on CampNet are conducted. Finally, the ability of few-shot learning and one-shot learning based on CampNet is explored.

\vspace{-0.2cm}
\subsection{Comparison between CampNet and Some other Method}
This section compares the performance of our proposed \textbf{CampNet} with three other speech editing methods, including \textbf{Synth}, \textbf{VoCo},  \textbf{Edit} by objective and subjective evaluations.

%

First, the objective results on the test sets of VCTK and LibriTTS are listed in Table \ref{table:a1}. In general, it can be found that the metrics on the two test datasets of CampNet are the best among all the systems. Specifically, in the frequency domain, the CampNet obtained the lowest MCD, which means that human perception would be better. Besides, the F0 has a significant influence on speech perception. We can find that CampNet achieves the best performance in F0-related metrics (F0-RMSE, V/UV error, and F0-CORR). The results show that CampNet can obtain more accurate fundamental frequency information.

Second, we show the waveforms and spectrograms of natural speech and the edited speech generated by different methods. We take the ${p225\_007.wav}$ in the VCTK corpus as an example, as shown in Fig. \ref{fig:figure2}. The region marked with time ($0.24s \sim 0.54s$) is the edited region. It is worth mentioning that the speaker $p225$ in the test set does not appear in the training set. It can be found that there will be an unnatural connection in the speech edited by \textbf{Synth} and \textbf{VoCo} models.
 There are apparent F0 discontinuities in the frequency domain. In addition, we draw the curve of F0. It can be found that the F0 of speech synthesized by \textbf{Synth} and \textbf{VoCo} are higher than the original region, while the speech synthesized by CampNet is consistent with the surrounding F0.


\begin{figure}[tp]
    \centering 
    \includegraphics[width=6.5cm]{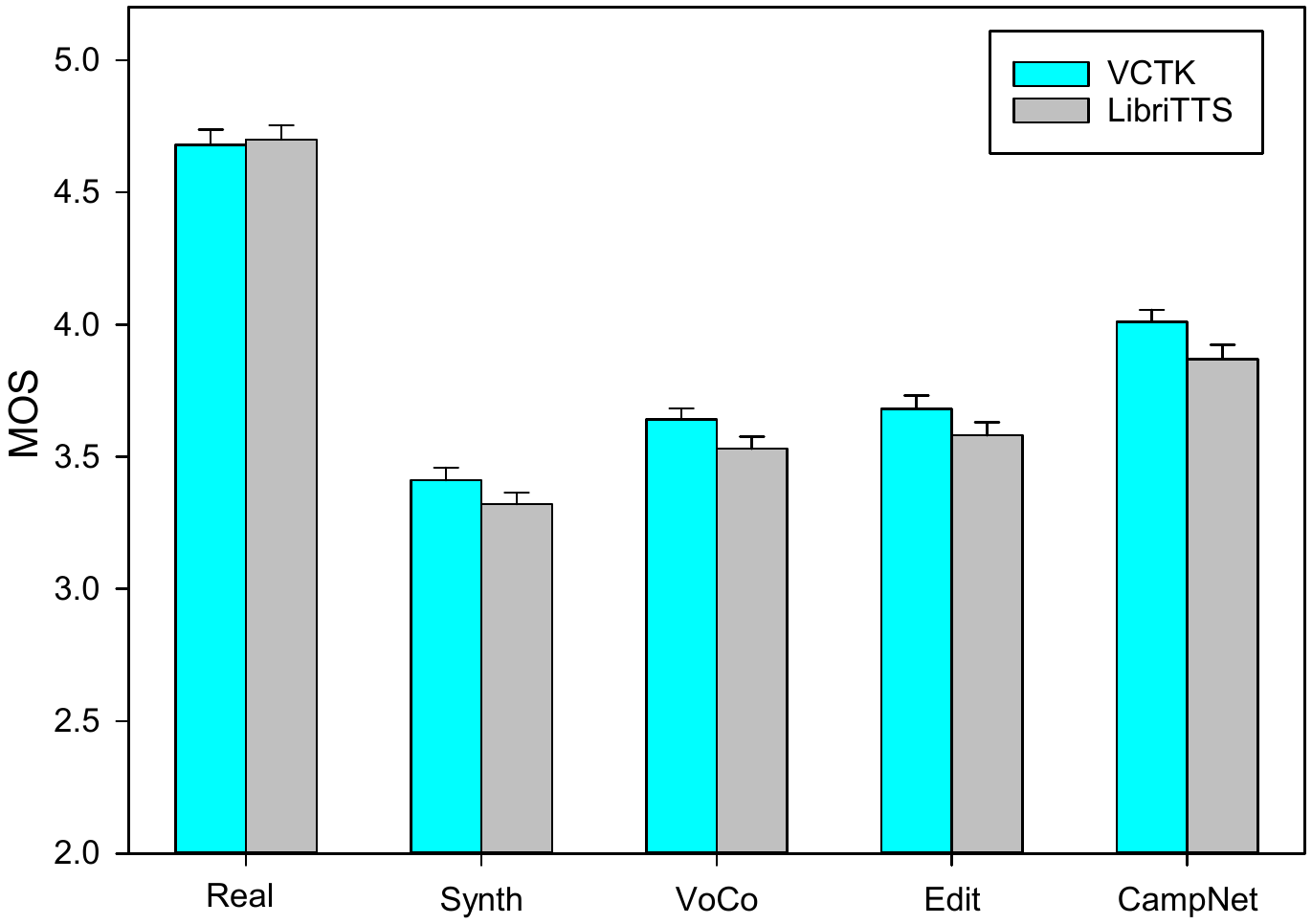}

    \caption{The MOS score with 95\% confidence intervals of the four systems and real speech.}
    \label{fig:mos}
\vspace{-0.5cm}
\end{figure}

\begin{figure}[tp]
    \centering 
    \includegraphics[width=6.5cm]{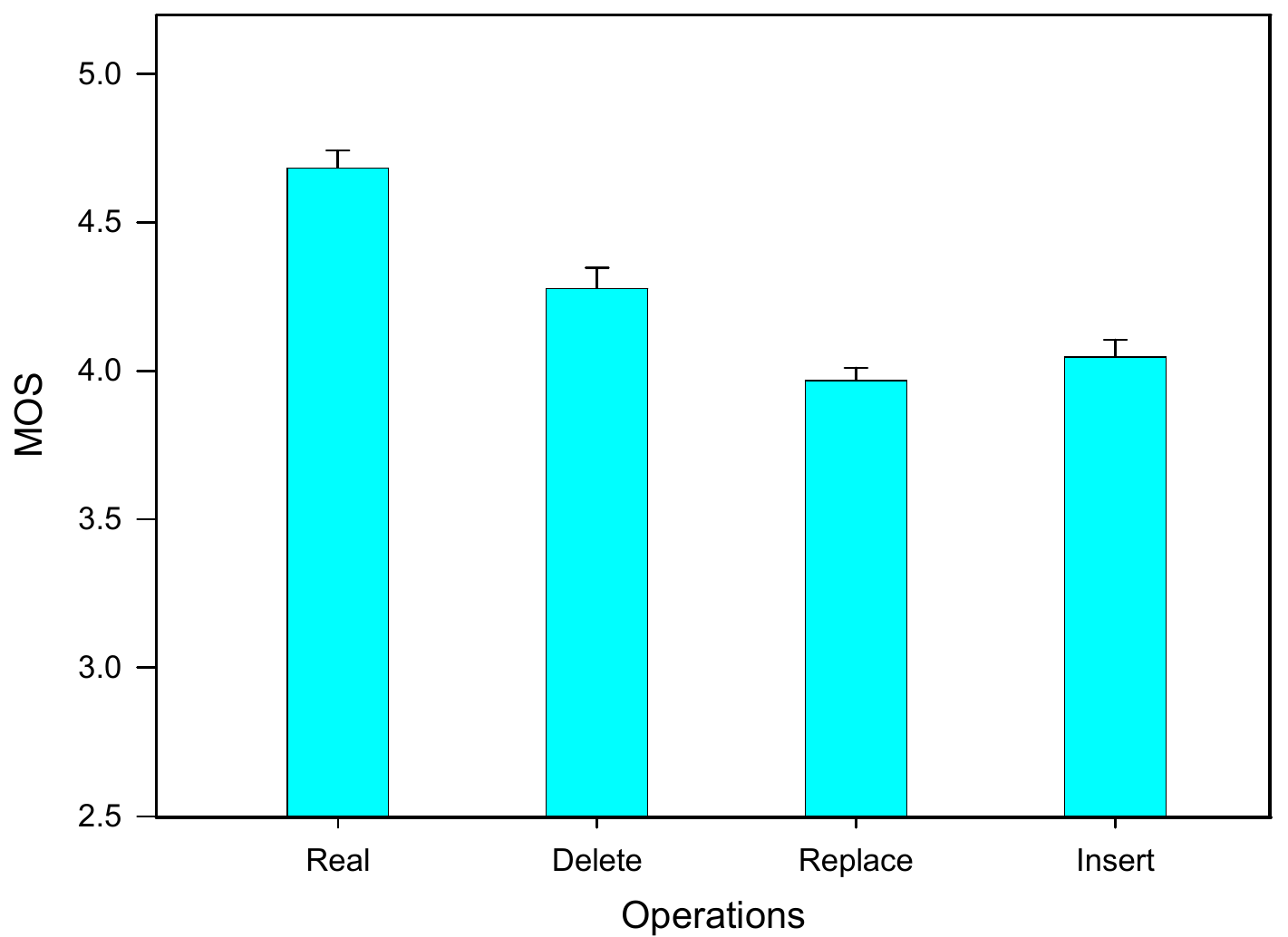}

    \caption{The MOS score with 95\% confidence intervals of different operations and real speech.}
    \label{fig:operation_mos}
\vspace{-0.5cm}
\end{figure}

Third, subjective evaluations are conducted to compare the performance of CampNet with other systems in terms of the naturalness of edited speech. In this evaluation, twenty utterances in each test set are selected and edited using the proposed method and other systems, including Synth, VoCo, and Edit. It should be noted that all the speakers of the test data did not appear in the training set. Twenty listeners took part in the evaluation. They are told in advance which word is predicted. The listeners were asked to listen and rate the quality of the restored sentence on a Likert scale \cite{joshi2015likert}: 1 = bad (very annoying), 2 = poor (annoying), 3 = fair (slightly annoying), 4 = good (perceptible but not annoying) and 5 = excellent (imperceptible, almost real). They can play the recording multiple times. Fig. \ref{fig:mos} shows the MOS score of each system. The results show that the CampNet is better than the other three systems in each test set. This is also consistent with the previous analysis of objective metrics. In addition, we compare the MOS scores of different operations. Specifically, we have prepared three test sets for the three operations. The speech of each test set has 20 sentences. Twenty listeners took part in the evaluation. They can play the recording multiple times. Fig.\ref{fig:operation_mos} shows the MOS score of each operation.  The results show that the MOS of delete operation is higher. The MOS score of insertion operation and replacement operation is similar, and their gap is small. 

In addition, we also compared CampNet with EditSpeech \cite{tan2021editspeech} and context-aware prosody correction method  \cite{morrison2021context}. To make the comparison more obvious, we directly  use the speech in the demo page of these systems as the standard, and use CampNet to generate the corresponding parallel speech as the comparison. We have put the samples on our demo page (\href{https://hairuo55.github.io/CampNet}{https://hairuo55.github.io/CampNet}), and we are looking forward to readers' listening. 

\subsection{Inference Speed}

We evaluate the inference speed of our proposed method with other systems. Since in the Synth and VoCo systems, Tacotron is used as the acoustic model, which significantly impacts the operation efficiency of the whole system. In this section, we compare CampNet with Tacotron and Transformer-TTS. The details of these TTS models are as follows:
 \begin{itemize}
\item \textbf{Tacotron} represents for the TTS model in which the decoder is based on LSTM. The structure details are the same as the acoustic model in the Synth system.
\item \textbf{Transformer-TTS} represents for TTS model which is based on transformer structure \cite{li2019neural}. The structure details are the same as the model in the paper \cite{li2019neural}. The number of encoder's blocks is 3, and the number of decoder's blocks is 9, which is consistent with the number of decoder blocks of CampNet. The number of hidden layer features is 256.
 \end{itemize}

The evaluation experiments are conducted with 52 Intel Xcon CPU, 512GB memory, and 1 NVIDIA V100 GPU. It is worth mentioning that, during the model's design, we have kept the model parameters as consistent as possible to eliminate their effects. Each model outputs 500 frames of acoustic features for fair comparison. We show the generation speed of acoustic features  in Table \ref{tab:speed}. It can be seen that the CampNet speeds up acoustic features generation by 34 times, compared with the Tacotron model. CampNet speeds up acoustic features generation by 215 times, compared with the Transformer-TTS model. It shows that autoregressive generation greatly affects the model's speed. CampNet can synthesize speech in the form of non-autoregressive and effectively improve the synthesis efficiency.

\begin{table}[t]
  \caption{The comparison of inference speed with 95\% confidence intervals for CampNet, Tacotron2 and Transformer-TTS. The value of inference speed indicates how long it takes to synthesize 500 frames of acoustic features.}
  \centering
  \begin{tabular}{c|c|c|c}
\hline \hline Model & Params & Inference(s) & Speedup \\
 \hline
    Tacotron &   3.94e7  & $1.501 \pm 0.280$ & $/$ ~~~  \\
    CampNet  &  \textbf{1.47e7} &  \textbf{0.044 $\pm$ 0.015} & $34\times$ ~~~ \\
 \hline
    Transformer-TTS  &  1.52e7  & $9.460 \pm 1.092$ & $/$ ~~~ \\
    CampNet  &  \textbf{1.47e7} & \textbf{0.044 $\pm$ 0.015} & $215\times$ ~~~ \\
 \hline  \hline
  \end{tabular}
  \label{tab:speed}
\vspace{-0.2cm}
  \end{table}

\begin{figure*}[tph]
    \centering 
    \includegraphics[width=15cm]{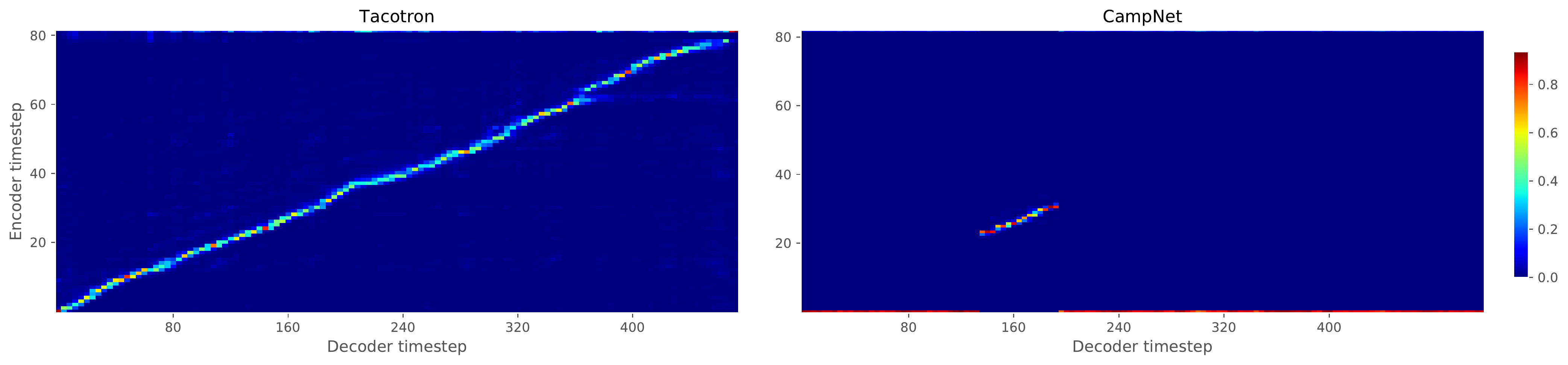}
\vspace{-0.1cm}
    \caption{Comparison of Tacotron alignment and CampNet alignment. The alignment of Tacotron is to align with the text in a complete time step. CampNet only aligns the edited speech region with the edited text.}
    \label{fig:figure1}
\vspace{-0.4cm}
\end{figure*}

\subsection{The Comparison of Alignment with Tacotron}\label{sec:ndps-com}

In this section, we explore the alignment mechanism of the CampNet model, which can help us understand the process of speech editing better. Since the coarse decoder is used to align text and speech, this section explores its attention mechanism.

First, we visualize the alignments of text and speech in the multi-head attention of coarse decoder, and the alignment of local sensitive attention in Tacotron is also visualized for comparison, as shown in Fig. \ref{fig:figure1}. We can find that the alignment of Tacotron is aligned in the whole time step, where each column denotes the attention probabilities corresponding to different encoder states for one decoder step. On the contrary, the alignment of CampNet is only in a small region, that is, the edited region. Its mapping area on the encoder is the edited text. Therefore, CampNet's attention mechanism only focuses on the edited region and automatically finds the text corresponding to the edited region. Compared with the whole alignment in Tacotron, the advantage of this method is that it can make the model pay more attention to the edited region and make the predicted speech consistent with the context.
Aligning only a region is simpler than aligning all, which is also in line with our intuitive understanding. Moreover, because text-based speech editing only requires partial alignment, the quality of the training corpus needed is not too high, while the TTS corpus should be as standard as possible.

\begin{figure}[t]
    \centering 
    \includegraphics[width=6.5cm]{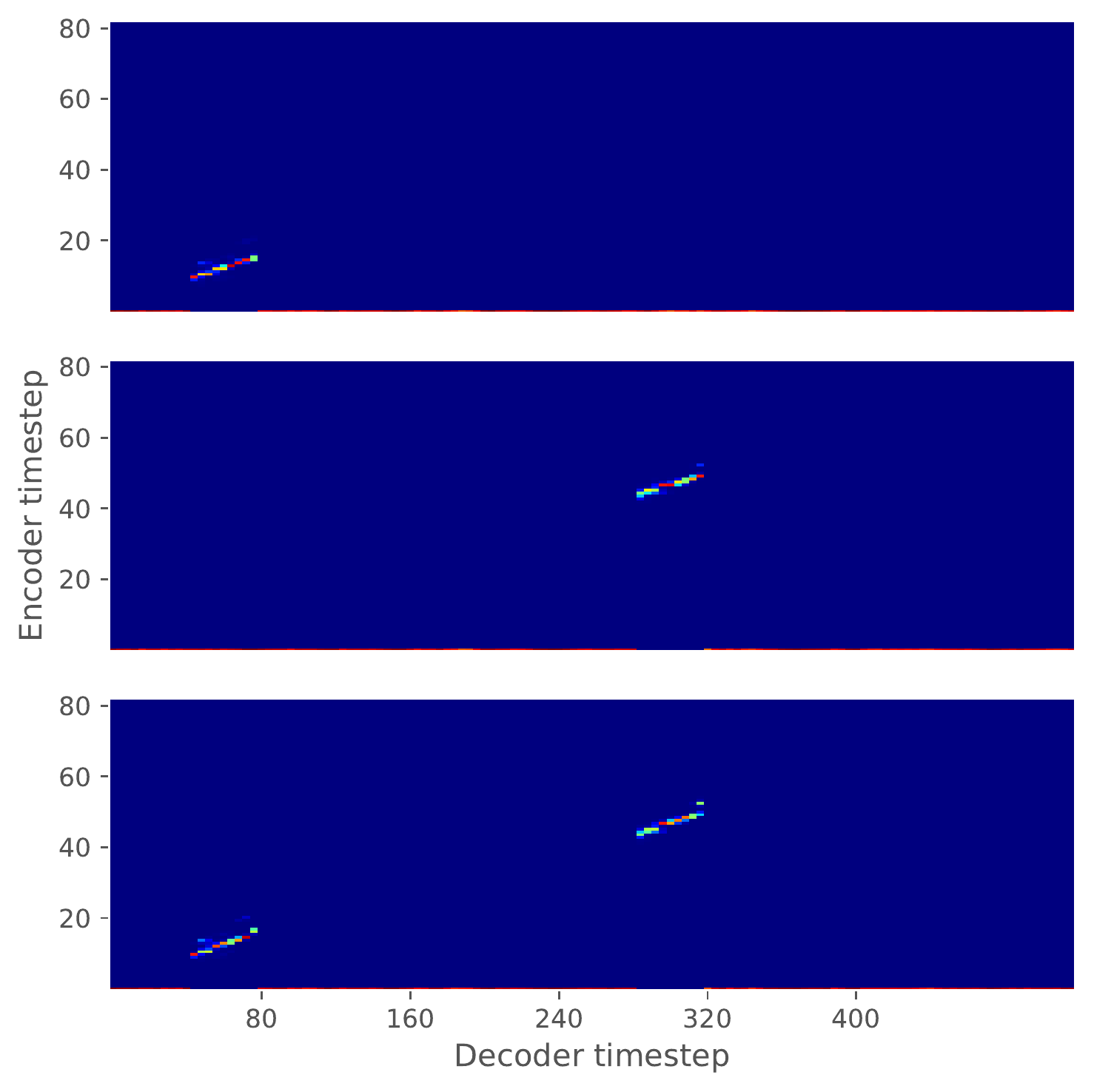}
    \caption{Alignments after masking the speech segments  at different positions. The last one is the alignment of  masking the two region at the same time.}
    \label{fig:ali3}
\vspace{-0.3cm}
\end{figure}

Fig. \ref{fig:ali3} shows the alignment results after masking different words in a sentence. It can be seen that the model has good robustness to edit different words in different positions. In addition, the last figure of Fig. \ref{fig:ali3} shows the alignment of two words edited simultaneously during the inference stage. Although we only mask one region of speech at the training stage, we can modify two different regions at the same time at the inference stage. This also verifies that the model can effectively learn the alignment between speech and text and only focus on the edited region.

\subsection{Effectiveness of Coarse-to-Fine Decoding}
As introduced in Section \ref{sec:decoder}, a coarse-to-fine decoding method is proposed to boost performance. To further understand the role of coarse-to-fine decoding, we compare the following three kinds of speech.

 \begin{itemize}
\item \textbf{One-decoder} represents the speech output by the model, which removed the coarse-to-fine decoding of CampNet. We directly use one decoder to output the final speech. To ensure the block number of the decoder are equal to the CampNet, we set the block numbers of the decoder as 9, which is the sum of the coarse decoder and fine decoder.
\item \textbf{Coarse-decoder} represents the speech $y^\mathrm{coaser}$ output by the coarse decoder of CampNet.
\item \textbf{Fine-decoder} represents the speech $y^\mathrm{fine}$ output by the fine decoder of CampNet.
 \end{itemize}

\begin{figure}[tp]
    \centering 
    \includegraphics[width=8cm]{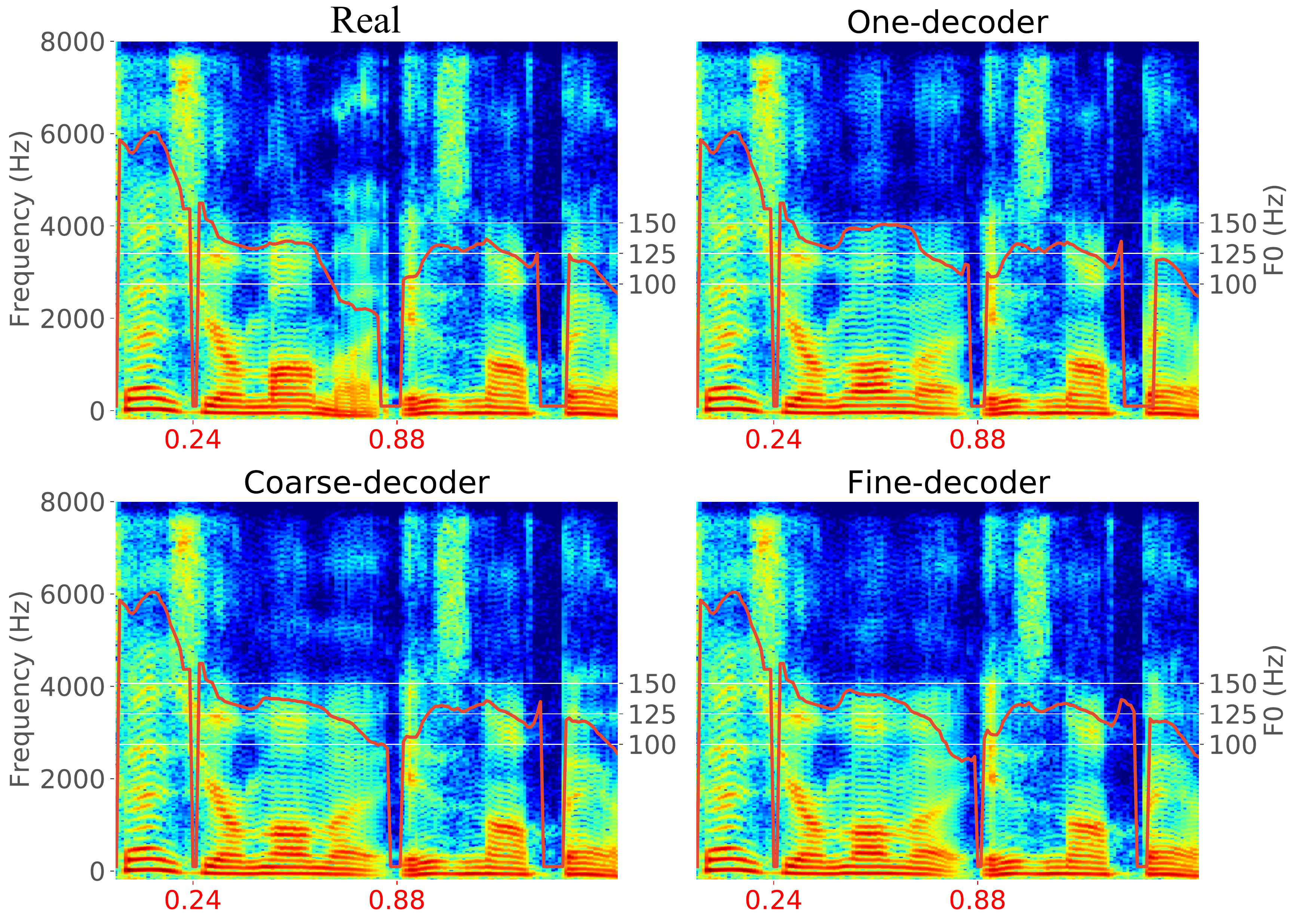}
    \caption{The spectrograms of the speech generated by different methods. The region marked with time ($0.24s \sim 0.88s$) is the edited region.}
    \label{fig:c_to_f}
\vspace{-0.3cm}
\end{figure}

First, we visualize some spectrograms of original speech and the edited speech after modifying one of the words, as shown in Fig. \ref{fig:c_to_f}, where 0.24s to 0.88s is the edited region. The speaker of the sample does not appear in the training corpus. From the perspective of F0, it can be found that the speech output by the fine decoder is closest to the F0 of the real speech. The F0 of the One-decoder is higher than the original F0. Although the F0 of the Coarse-decoder is different from the real one, after adjusting by the fine decoder, a similar F0 curve is obtained. Besides, we can find that the speech of the fine decoder has a more precise spectrum.

\begin{table}[t]
    \centering
        \caption{OBJECTIVE EVALUATION RESULTS OF  ONE-DECODER, COARSE-DECODER AND
FINE-DECODER ON THE TEST SETS OF VCTK AND LIBRITTS.}
\scalebox{0.92}{
\begin{tabular}{cc|ccc}

\hline \hline  &   & Coaser-decoder & One-decoder & Fine-decoder  \\
\hline &  MCD(dB)  & 0.395   & 0.388    &\textbf{0.380} \\
VCTK & F0-RMSE(Hz)    &   9.108  & 8.851   & \textbf{8.637} \\
  & V/UV error(\%)    & 1.712 &  1.686      & \textbf{1.635} \\ 
& F0-CORR   &0.977 &       0.978     & \textbf{0.981}   \\

\hline &  MCD(dB)  &0.639  & 0.632    &\textbf{0.628} \\
LibriTTS & F0-RMSE(Hz)    & 21.650   & 20.859   & \textbf{20.201} \\
  & V/UV error(\%)    & 3.859 & 3.766       & \textbf{3.675} \\ 
& F0-CORR   & 0.945 & 0.951    & \textbf{0.954}   \\
\hline \hline
\end{tabular}}
\label{table:c-to-f}
\vspace{-0.2cm}
\end{table}

\begin{table*}[t]
    \centering
        \caption{OBJECTIVE EVALUATION RESULTS OF  DIFFERENT MASK RATIO AT INFERENCE STAGE ON THE  VCTK TEST SET}
\scalebox{1}{
\begin{tabular}{c|cccccc|cccccc}
\hline \hline    &  & &  \multicolumn{2}{c}{VCTK}   & & & & & \multicolumn{2}{c}{LibriTTS}    & &  \\
 \hline Metrics & M-6\%  & M-8\%   & M-10\%  & M-12\%  & M-14\% & M-16\%  & M-6\%  & M-8\%   & M-10\%  & M-12\%  & M-14\% & M-16\% \\
\hline  MCD(dB)  & 0.465  &  0.387  & 0.391   & \textbf{0.380}  &  0.383 & 0.398  & 0.746  & 0.661   & 0.631 & \textbf{0.628}  & 0.634 & 0.650 \\
F0-RMSE(dB)    & 10.511    & 9.723  & 9.407 &  \textbf{8.637}  & 9.255   &9.114 & 22.895  &22.242 & 21.049  & \textbf{20.201}   & 21.086 & 21.820 \\
   V/UV error  & 1.989 &   1.658      & 1.610 & 1.635  &\textbf{1.492}     &1.750   & 5.679  & 4.136   & \textbf{3.635}  & 3.675  & 4.000 & 4.259\\ 
 F0-CORR  & 0.971          &0.976   &0.977 &  \textbf{0.981}  &  0.978   &0.978 & 0.952  & 0.943   & 0.949  & \textbf{0.954}   &  0.948 &  0.945\\
\hline \hline
\end{tabular}}
\label{table:train_change}
\vspace{-0.2cm}
\end{table*}



Second, we compare the objective metrics of the three kinds of speech, as shown in Table \ref{table:c-to-f}. It can be found that the speech of the fine decoder achieves the best performance in all metrics on the two test sets. Specifically, the speech output by the fine decoder is closer to the natural speech in both spectrum and F0, which shows that coarse-to-fine decoding can learn more accurate frequency domain information.

Third, we conduct a subjective ABX test to compare the three kinds of speech. In each subjective test, twenty sentences are randomly selected from the LibriTTS test set. Twenty listeners evaluate each pair of generated speech. The listeners are asked to judge which utterance in each pair has better speech quality or no preference in the edited area. They were told in advance which word was predicted. 
The results are listed in Table \ref{table:abx1}. Fine-decoder outperformes Coarse-decoder and One-decoder, which is consistent with the objective metrics analysis. This result further shows  the effectiveness of coarse-to-fine decoding. 

It is worth noting that whether the model is improved can be judged through numerical comparison of the objective metrics, but this is not absolutely relevant. More importantly, we should judge it by subjective evaluation. We have put the relevant samples on the demo page (https://hairuo55.github.io/CampNet/), and we recommend readers to listen.

\begin{table}[t]
\centering
\caption{AVERAGE PERFERENCE SCORE(\%) ON SPEECH QUALITY AMONG DIFFERENT SYSTEMS, WHERE N/P STANDS FOR "NO PERFERENCE". THE $p$-VALUES \textless 0.01.}
\scalebox{1}{
\begin{tabular}{c|ccc|c}
\hline
\hline
  System & Scores & N/P & Scores & System \\
 A  & \textbf{A(\%)} & Neural(\%) & B(\%)  & B \\ \hline
CoareDecoder    &  22.25   & 47.00 & 30.75 & OneDecoder   \\
FineDecoder  & 43.25   &  32.75  & 24.00 &OneDecoder    \\
FineDecoder  &44.75   & 35.00 & 20.25  &CoareDecoder   \\ \hline \hline
\end{tabular}}
\label{table:abx1}
\end{table}


\subsection{The Ability of Editing Different Length}

\begin{figure}[t]
    \centering 
    \includegraphics[width=7.5cm]{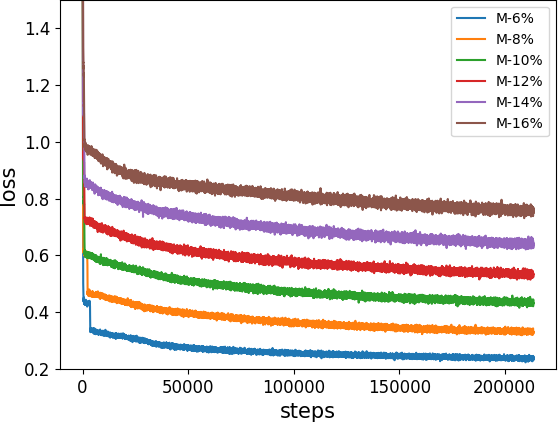}
    \caption{Comparison of loss functions with different mask ratios at the training stage. The smaller the masked region is, the easier the model is to be optimized and the lower the loss is.}
    \label{fig:aa_diffmask}
\vspace{-0.3cm}
\end{figure}

\begin{table}[t]
    \centering
        \caption{OBJECTIVE EVALUATION RESULTS OF  DIFFERENT MASK RATIO AT INFERENCE STAGE ON THE  VCTK TEST SET}
\scalebox{1}{
\begin{tabular}{c|ccccc}
\hline \hline   & 0.5s  & 1.0s   & 1.5s  & 2.0s & 2.5s  \\
\hline  MCD(dB)  & \textbf{0.345}  &  0.628  & 0.960   & 1.296  &  1.387  \\
F0-RMSE(dB)    & \textbf{4.577}     & 7.133 & 10.887 &   13.403  &  14.584   \\
   V/UV error  & \textbf{1.489}  &   2.783      & 4.626 &  6.201  & 8.175    \\ 
 F0-CORR  & \textbf{0.994}    & 0.986    & 0.964 & 0.945  &  0.931 \\
\hline \hline
\end{tabular}}
\label{table:test_change}
\vspace{-0.4cm}
\end{table}


As introduced in Section \ref{sec:few-shot}, we propose the mask prediction method to simulate the speech editing process. The core idea is to randomly mask a speech region at the training stage and then predict the masking region. There may be a mismatch between the length of the masking region at the training stage and the inference stage, which may affect the model's performance. This section mainly explores the impact of this problem on the model performance.

First, we explore the influence of different mask ratios during training. Specifically, the mask ratios at the training stage are set to 6\%, 8\%, 10\%, 12\%, 14\%, and 16\%, respectively. The trained models are represented by M-6\%, M-8\%, M-10\%, M-12\%, M-14\%, and M-16\%. All models are trained for 2 million steps with the same structure and hyper-parameters. Fig. \ref{fig:aa_diffmask} compares the change of loss of each model with the increase of training steps. Obviously, with the rise of mask ratios, the loss function becomes larger and larger, which means it is more difficult for the model to predict the masked speech. The smaller the mask ratios, the easier the model is to be optimized.
Further, to test the effect of each model, we calculate the objective metrics of each model on the test set, which is shown in Table \ref{table:train_change}. It can be found that when the mask ratio is set to 12\%, the model has the best effect on most indicators (MCD, F0-RMSE, and F0-CORR). Therefore, we use a mask ratio of 12\% in the training phase, which performs well in the test set and can also optimize the training loss to a suitable level.

Second, we explore the impact of the edited region's length at the inference stage in one-step.
In order to control the length of the editing area in the test set within the specified length, we directly mask a region in a fixed length and use CampNet to predict the mask region. Referring to Table \ref{table:train_change}, we set the mask ratio as 12\% during the training stage. For all speech in the test set, we take a fixed position of the speech as the starting point, then mask the speech with different lengths, which are 0.5s, 1.0s, 1.5s, 2.0s, and 2.5s, respectively. We input the masked speech into CampNet to predict the masked speech.
We calculate the objective metrics of the predicted speech and the real speech, as shown in Table \ref{table:test_change}. It can be  found that with the increase of mask length on the test set, the model's performance shows a downward trend. The best effect on the test set is the smallest mask length, that is \textbf{0.5s}. This indicates that the smaller the mask area is, the closer the synthesized speech is to the real speech. When the mask length exceeds 1.5s, the synthesized speech is quite different from the actual speech in the sense of hearing. Therefore, the length of the editing area should not be too long during the inference stage, preferably less than 1.5s under the condition that the mask ratio at the training stage is 12\%. It is worth mentioning that this length is sufficient to deal with common scenes in text-based speech editing, such as replacing wrong pronunciation or inserting emphasis words.
\vspace{-0.2cm}


\begin{figure}[t]
    \centering 
    \includegraphics[width=8.5cm]{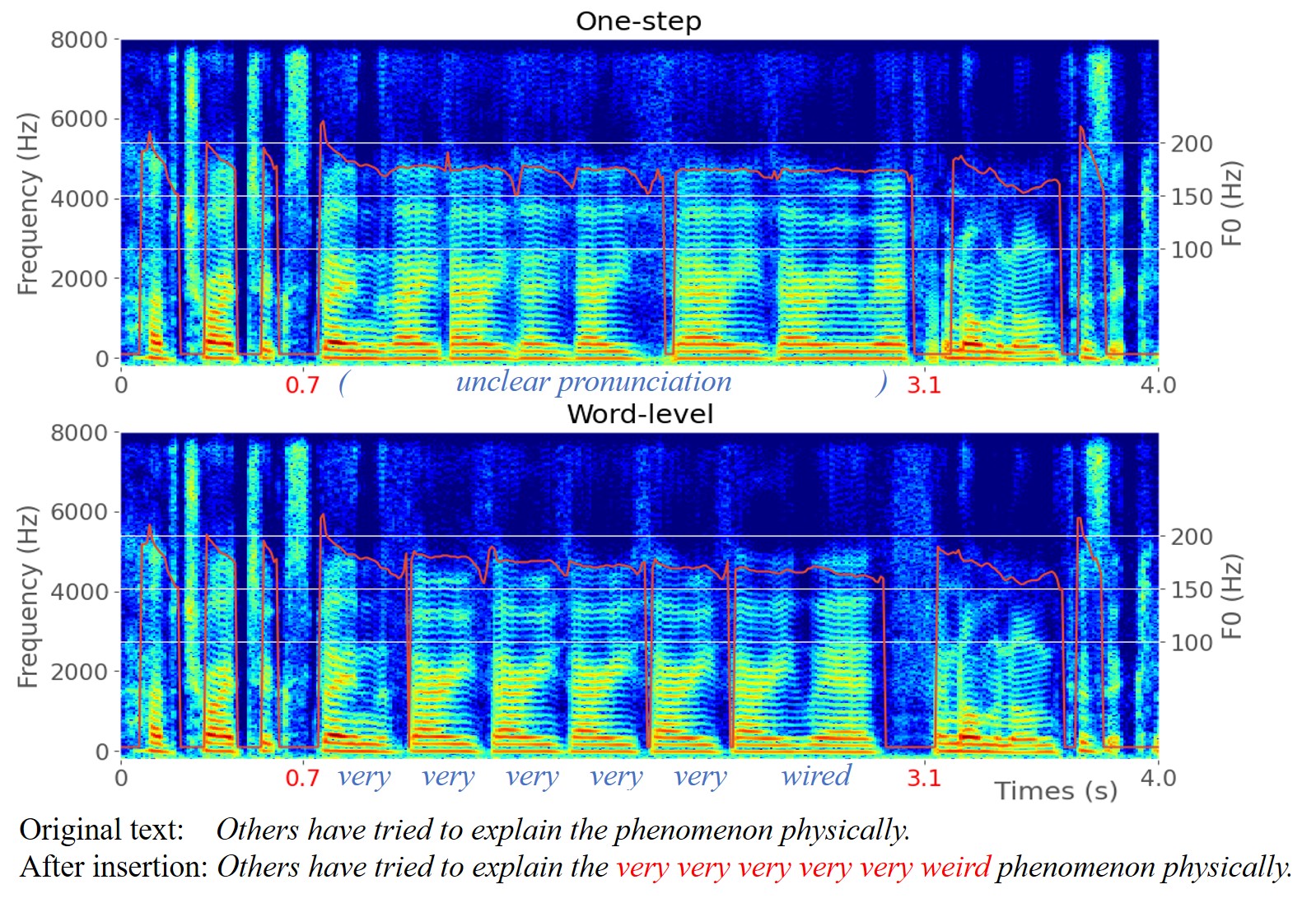}
    \caption{The spectrograms of the speech generated by different methods. The region marked with red color ($0.7s \sim 3.1s$) is the insertion area, and its correspondin text is \emph{"very very very very very weird"}. The pronunciation in the insertion area of method One-step is not clear, while the pronunciation of method Word-level is natural and normal.}
    \label{fig:word-level-exp}
\vspace{-0.3cm}
\end{figure}

\subsection{Effectiveness of Word-level Autoregressive Generation Method}
As introduced in Section \ref{sec:word-level}, we propose a word-level autoregressive generation method based on CampNet to face the situation of generating speech with long text. In this section, to explore the ability of word-level autoregressive generation method,  we take the insert operation as an example, different from the previous experiments, the length of the words to be replaced is much longer than that used in the previous experiments. The number of words to be inserted is more than 5, and the corresponding speech length is more than 2 seconds. We compared the following two generation methods:
\begin{itemize}
     \item \textbf{One-step} means to generate the speech of all words to be inserted at one step.
     \item \textbf{Word-level} means to generate the speech by using word-level autoregressive generation method, which is introduced in Section \ref{sec:word-level}.
 \end{itemize}

First, we compare the spectrum generated by the two methods after inserting some words, as shown in Fig. \ref{fig:word-level-exp}.  It can be seen that when the word-level autoregressive generation method is adopted,  the generated speech is more stable. When multiple words are generated at one step, the pronunciation of the synthesized speech will be unclear. The reason is that  only short speech segments is masked in the training stage, which makes it impossible to generate long segments of speech during the inference stage. While, the word-level autoregressive generation method can ensure the matching of mask length in inference  and training stage.

\begin{table}[tph]
\centering
\caption{AVERAGE PERFERENCE SCORE(\%) ON SPEECH QUALITY AMONG DIFFERENT SYSTEMS, WHERE N/P STANDS FOR "NO PERFERENCE". THE $p$-VALUES \textless 0.01.}
\scalebox{1}{
\begin{tabular}{c|ccc|c}
\hline
\hline
  System & Scores & N/P & Scores & System \\
 A  & \textbf{A(\%)} & Neural(\%) & B(\%)  & B \\ \hline
One-step    &  5.25   & 8.00 & 86.75 & Word-level   \\
 \hline \hline
\end{tabular}}
\label{table:word-level-abx}

\end{table}

Second, since there is no real speech as a comparison when inserting long text, it is not convenient for us to make objective evaluation. Therefore, we conduct a subjective ABX test to compare the two methods. In each subjective test, twenty sentences are randomly selected. Twenty listeners evaluate each pair of generated speech. The listeners are asked to judge which utterance in each pair has better speech quality or no preference in the edited area. They were told in advance which word was predicted. 
The results are listed in Table \ref{table:word-level-abx}.  It is obviously that, in the case of a large number of words to be inserted, the generation method based on word-level autoregression is better than the one-step generation method. For more samples, please refer to our demo page.

\begin{table*}[tph]
    \centering
        \caption{OBJECTIVE EVALUATION RESULTS OF  DIFFERENT METHODS ON THE TEST SETS.}
\scalebox{1}{
\begin{tabular}{cc|cccc}

\hline \hline Speaker & Metrics & 1-utt\_wo\_finetune & 1-utt\_w\_finetune & 50-utts\_w\_finetune   \\
\hline & MCD(dB)  &0.395  &\textbf{0.273}    & 0.275   \\
P225 & F0-RMSE(Hz)    & 7.557   & 4.630   & \textbf{4.614} \\
(female) &  V/UV error(\%)    & 1.655  &  \textbf{1.325}       & 1.338   \\ 
& F0-CORR   & 0.983 & \textbf{0.992} & \textbf{0.992}   \\
\hline
& MCD(dB)  & 0.367  &0.272    &\textbf{0.271} \\
P226 & F0-RMSE(Hz)    & 9.406  &7.740   & \textbf{6.987} \\
(male) &  V/UV error(\%)    & 1.627  &  1.173       & \textbf{1.137} \\ 
& F0-CORR   & 0.958 &0.966    & \textbf{0.978}   \\
\hline \hline
\end{tabular}}
\label{table:fewshot}

\end{table*}

 \begin{table*}[tph]
 \centering
 \caption{AVERAGE PREFERENCE SCORES (\%) ON SPEECH QUALITY OF DIFFERENT  METHODS, WHERE N/P STANDS FOR “NO
PREFERENCE”, AND $p$ DENOTES THE $p$-VALUE OF A $t$-TEST
}
 \begin{tabular}{c|ccccc}
 \hline
 \hline
  &  1-utt\_wo\_finetune & 1-utt\_w\_finetune & 50-utts\_w\_finetune & N/P & p \\ \hline
  1-utt\_wo\_finetune vs  50-utts\_w\_finetune  & 11.75  & --  & 66.00 & 22.25 &   \textless  0.01 \\
 1-utt\_w\_finetune vs   50-utts\_w\_finetune  &  --   & 19.50 & 60.75  & 19.75 & \textless  0.01 \\
 1-utt\_wo\_finetune  vs 1-utt\_w\_finetune   & 27.25    & 36.00   & -- & 36.75 &  \textless  0.01 \\ \hline \hline
 \end{tabular}
 \label{table:fewoshotabx}
 \end{table*}

\subsection{The Ability of One-shot and Few-shot Learning}

As introduced in Section \ref{sec:shot}, we propose a transfer learning method based on CampNet. In this section, we explore the ability of CampNet to face few-shot and one-shot learning by comparing the following three models:
\begin{itemize}
     \item \textbf{1-utt\_wo\_finetune} means to directly edit the speech of an unseen speaker using CampNet without fine-tuning the model.
     \item \textbf{1-utt\_w\_finetune} means to fine-tune CampNet with one sentence from an unseen speaker before speech editing, which is introduced in Section \ref{sec:one-shot}.
     \item \textbf{50-utts\_w\_finetune} means to fine-tune CampNet with 50 sentences from an unseen speaker before speech editing, which is introduced in Section \ref{sec:few-shot}.
 \end{itemize}

The steps of fine-tuning are five epochs of the  dataset used to fine-tune. The fine-tuning datasets are separate from the data used to calculate objective and subjective  metrics.
We calculate the objective metrics of the synthesized speech and real speech of each system, as shown in Table \ref{table:fewshot}. We can find that fine-tuning the model with a small amount of corpus can significantly improve the performance in objective metrics. Even if one utterance is used for fine-tuning the model, it can be found that the objective metrics are improved by comparing system \textbf{1-utt\_w\_finetune} and \textbf{1-utt\_wo\_finetune}. This shows the effectiveness of using only one sentence to adapt the model.

Besides, we conduct an ABX test on the three methods. Twenty sentences of each speaker are synthesized by two comparative systems. Twenty listeners evaluate each pair of generated speech. The listeners are asked to judge which utterance in each pair has better speech quality or no preference. The $p$-value is used to measure the significance of the difference between two systems. The results are listed in Table \ref{table:fewoshotabx}. It can be found that the quality of speech can be further improved through fine-tuning. Furthermore, using only one sentence, the adaptive model is also significantly improved than the non-adaptive model.


\vspace{-0.1cm}
\section{Conclusion}\label{sec:5}
This paper has proposed a context-aware mask prediction network for the end-to-end text-based speech editing task, which can delete, replace and insert the speech at the word level by editing the transcription. To simulate the speech editing process at the training stage, the text-based speech editing task is viewed as a two-stage process: masking and prediction, and a coarse-to-fine decoding method is proposed to achieve context-aware prediction. At the inference stage, three operations are designed based on CampNet, corresponding to the deletion, insertion, and replacement operations. Then,   to synthesize  the speech of arbitrary length text in insertion and replacement operations, a word-level autoregressive generation method is proposed.
 Finally, we propose a one-sentence speaker adaptation method for the CampNet and explore the ability of few-shot and one-shot learning based on CampNet, which can boost performance further by using only one sentence. Compared with TTS and VC, it also provides a new method for speech forgery. The experimental results demonstrate that the CampNet is better than the TTS, VoCo, and manual editing in subjective evaluation, objective evaluation, and operational efficiency for the text-based speech editing task. In addition, the few-shot learning ability based on CampNet is better than TTS and VC systems.  Improving the speech quality further based on CampNet is the future work.

\section{Acknowledgment}
This work is supported by the National Key Research and Development Plan of China (No.2020AAA0140003), the National Natural Science Foundation of China (NSFC) (No.62101553, No.61901473, No.61831022), the Key Research Project (No.2019KD0AD01),  and  is also  partially funded by Huawei Noah's Ark Lab.


%





\ifCLASSOPTIONcaptionsoff
  \newpage
\fi



%



\small
\bibliographystyle{IEEEtran}
\bibliography{refs}

%




%
\begin{IEEEbiography}[{\includegraphics[width=1.1in,height=1.25in,clip,keepaspectratio]{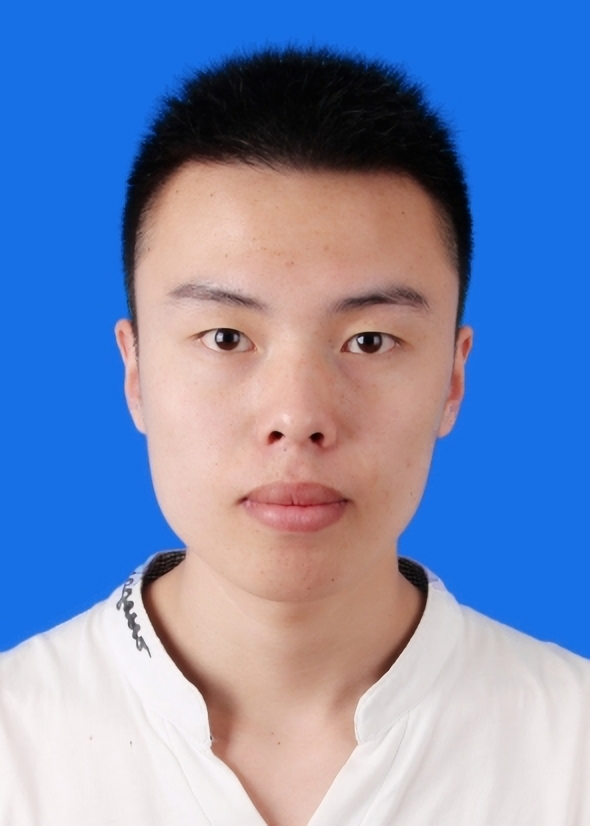}}]{Tao Wang}
received the B.E. degree from the
Department of  Control Science and Engineering, Shandong University (SDU), Jinan, China, in 2018. He is currently working
toward the Ph.D. degree with the National Laboratory
of Pattern Recognition, Institute of Automation (NLPR), Chinese
Academy of Sciences (CASIA), Beijing, China. His current
research interests include speech synthesis, voice conversion,  machine learning, and transfer learning.
\end{IEEEbiography}
\vspace{-1.0cm}

\begin{IEEEbiography}[{\includegraphics[width=1.1in,height=1.25in,clip,keepaspectratio]{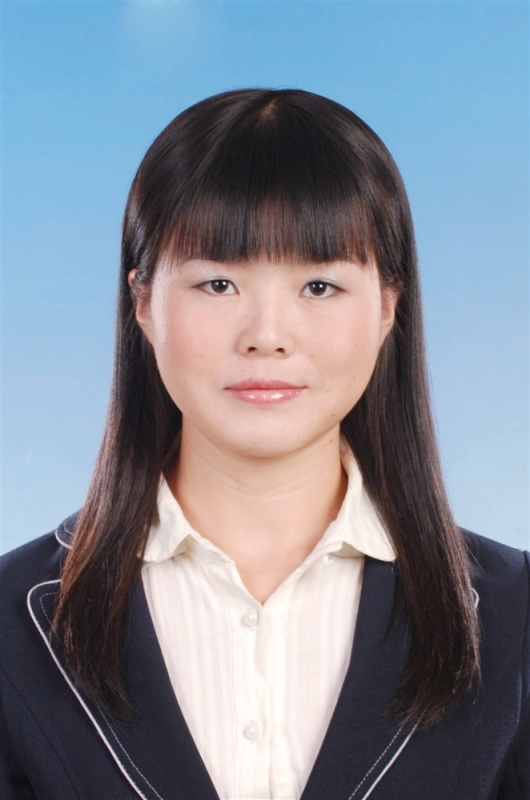}}]{Jiangyan Yi}
received the Ph.D. degree from the
University of Chinese Academy of Sciences, Beijing,
China, in 2018, and the M.A. degree from the Graduate
School of Chinese Academy of Social Sciences,
Beijing, China, in 2010. She was a Senior R\&D
Engineer with Alibaba Group from 2011 to 2014.
She is currently an Assistant Professor with the National
Laboratory of Pattern Recognition, Institute of
Automation, Chinese Academy of Sciences, Beijing,
China. Her current research interests include speech
processing, speech recognition, distributed computing,
deep learning, and transfer learning.
\end{IEEEbiography}
\vspace{-1.0cm}

\begin{IEEEbiography}[{\includegraphics[width=1.1in,height=1.25in,clip,keepaspectratio]{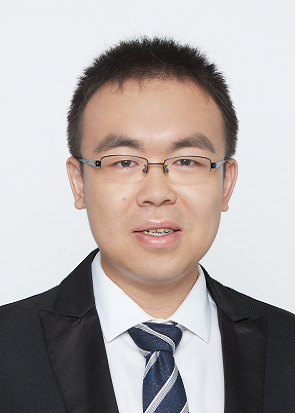}}]{Ruibo Fu}
is an assistant professor in National
Laboratory of Pattern Recognition, Institute of Automation, Chinese Academy
of Sciences, Beijing. He obtained B.E. from Beijing University of Aeronautics and Astronautics in 2015 and Ph.D. from Institute of Automation, Chinese Academy of Sciences in 2020. His research interest is speech synthesis and transfer learning. He has published more than 10 papers in international conferences and journals such as ICASSP and INTERSPEECH and has won the best paper award twice in NCMMSC 2017 and 2019.
\end{IEEEbiography}
\vspace{-1.0cm}

\begin{IEEEbiography}[{\includegraphics[width=1.1in,height=1.25in,clip,keepaspectratio]{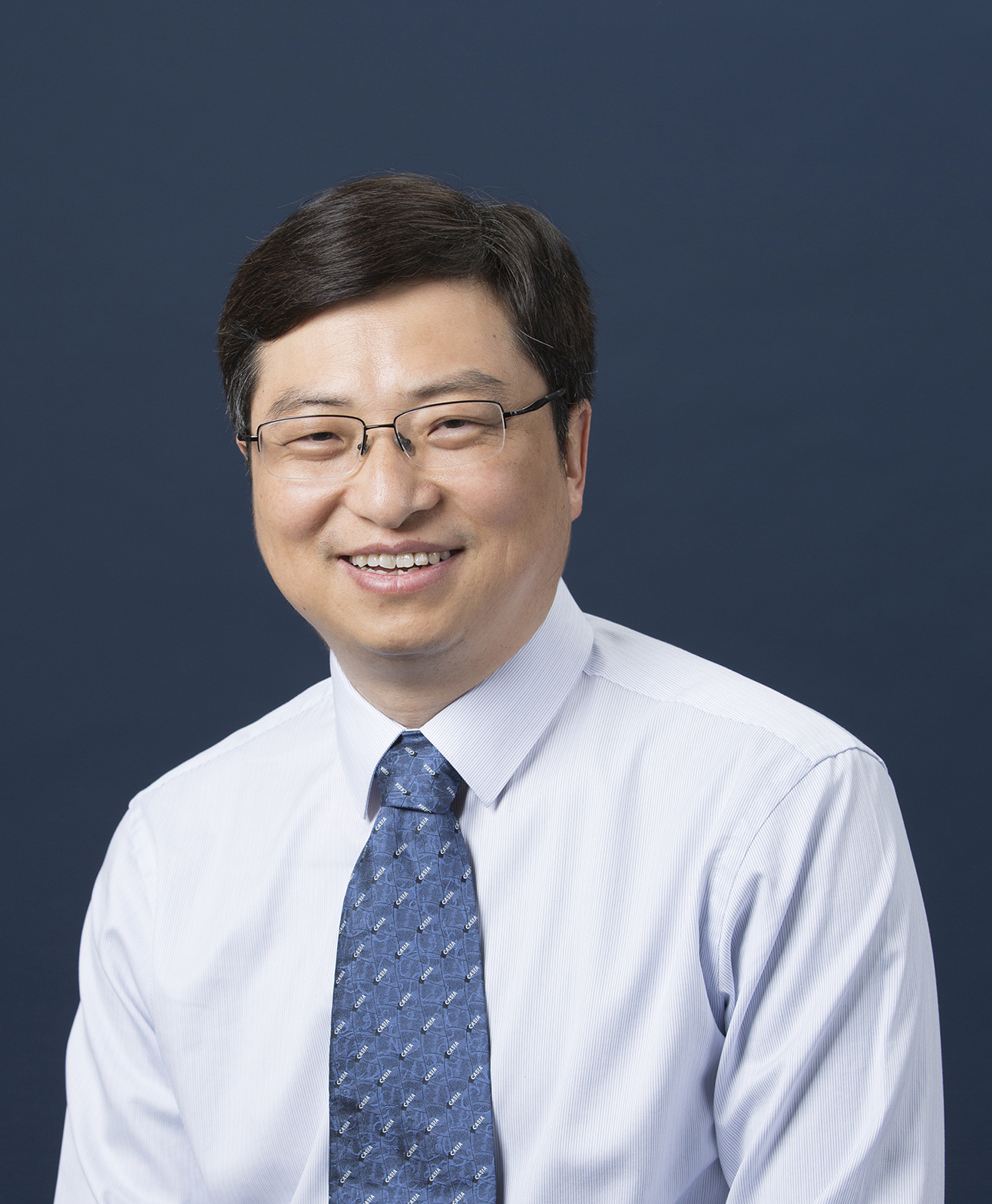}}]{Jianhua Tao}
(SM’10) received the Ph.D. degree from
Tsinghua University, Beijing, China, in 2001, and
the M.S. degree from Nanjing University, Nanjing,
China, in 1996. He is currently a Professor with
NLPR, Institute of Automation, Chinese Academy
of Sciences, Beijing, China. His current research interests
include speech synthesis and coding methods,
human computer interaction, multimedia information
processing, and pattern recognition. He has authored
or coauthored more than 80 papers on major journals
and proceedings including IEEE TRANSACTIONS ON
AUDIO, SPEECH, AND LANGUAGE PROCESSING, and received several awards
from the important conferences, such as Eurospeech, NCMMSC, etc. He serves
as the chair or program committee member for several major conferences,
including ICPR, ACII, ICMI, ISCSLP, NCMMSC, etc. He also serves as the
steering committee member for IEEE Transactions on Affective Computing, an
Associate Editor for Journal on Multimodal User Interface and International
Journal on Synthetic Emotions, the Deputy Editor-in-Chief for Chinese Journal
of Phonetics.
\end{IEEEbiography}
\vspace{-1.0cm}

\begin{IEEEbiography}[{\includegraphics[width=1.1in,height=1.25in,clip,keepaspectratio]{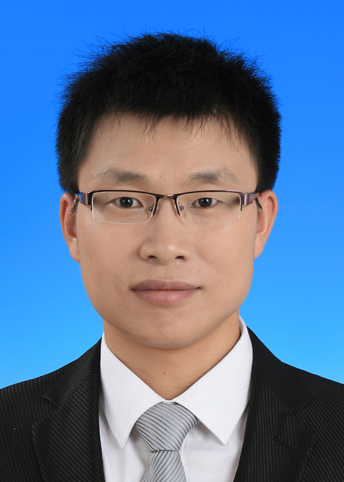}}]{Zhengqi Wen}
received the B.E. degree from the
Department of Automation, University of Science and
Technology of China, Hefei, China, in 2008 and the
Ph.D. degree from the National Laboratory of Pattern
Recognition, Institute of Automation, Chinese
Academy of Sciences, Beijing, China, in 2013. From
March 2009 to June 2009, he was an intern student
with Nokia Research Center, China. From December
2011 to March 2012, he was an intern student
with the Faculty of Systems Engineering,Wakayama
University, Japan. From July 2014 to January 2015,
he was a visiting scholar, under the supervision of Professor Chin-Hui Lee,
with the School of Electrical and Computer Engineering, Georgia Institute of
Technology, USA. He is currently an Associate Professor with the National
Laboratory of Pattern Recognition, Institute of Automation, Chinese Academy
of Sciences, Beijing, China.
\end{IEEEbiography}




\end{document}